\documentclass[%
 aip,
 amsmath,amssymb,
reprint,%
]{revtex4-1}
\usepackage{float} 
\usepackage{graphicx}
\usepackage{dcolumn}
\usepackage{bm}

\usepackage[utf8]{inputenc}
\usepackage[T1]{fontenc}
\usepackage{mathptmx}
\usepackage{etoolbox}

\makeatletter
\def\@email#1#2{%
 \endgroup
 \patchcmd{\titleblock@produce}
  {\frontmatter@RRAPformat}
  {\frontmatter@RRAPformat{\produce@RRAP{*#1\href{mailto:#2}{#2}}}\frontmatter@RRAPformat}
  {}{}
}%
\makeatother
\begin{document}

\preprint{AIP/123-QED}


\title{Temperature and mean axial momentum vs.\ laser intensity of electrons
released from O$_2$ by an 800~nm ultrashort pulsed laser}
\author{Edward L. Ruden}

\affiliation{Air Force Research Laboratory, Directed Energy Directorate}

\begin{abstract}
A semi-empirical model is presented for the thermalized temperature $T$ and
mean momentum in the direction of laser propagation $\left\langle p_{{%
\mathrm{f}z}}\right\rangle $ of electrons released from O$_{2}$ after the
passage of a focused $800$~nm ultrashort pulsed laser pulse vs.\ peak laser
intensity $I_{0}$ to provide initial conditions for plasma electrodynamic
simulations. For this, theoretical kinetic energy spectra based on the most
probable ionization path of a nonadiabatic Strong Field Approximation (SFA)
are fit to experimental spectra by two adjustable parameters assumed to
represent the effects of electron recombination with or rescatter off of its
parent ion during the optical cycle subsequent to ionization. These are
effects not treated by the original SFA. The classical kinematics of elastic
rescatter and the Lorentz force, then, are used to estimate $\left\langle p_{%
{\mathrm{f}z}}\right\rangle $.
\end{abstract}

\volumeyear{year}
\volumenumber{number}
\issuenumber{number}
\eid{identifier}
\date[Date text]{date}
\received[Received text]{date}

\revised[Revised text]{date}

\accepted[Accepted text]{date}

\published[Published text]{date}

\startpage{1}
\endpage{2}
\maketitle

\section{Introduction}

The purpose of this paper is to present a semi-empirical model of the
thermalized electron temperature $T$ and mean momentum in the direction of
laser propagation $\left\langle p_{{\mathrm{f}z}}\right\rangle $ of free
electrons released from O$_{2}$ vs.\ peak laser intensity $I_{0}$ after
passage of the pulse of a Ti:sapphire ultrashort pulsed laser (USPL) with
wavelength $\lambda =800$~nm. $p_{{\mathrm{f}z}}$ refers to the \emph{%
post-optical} electron momentum in the (laser) $z$-direction.
\textquotedblleft f\textquotedblright\ appears in the subscript of
properties at the time when the pulse has just \textquotedblleft
finished\textquotedblright , but thermalization has not had time to occur;
such properties are referred to as being \textquotedblleft
post-optical\textquotedblright . \textquotedblleft $\left\langle
{}\right\rangle $\textquotedblright\ denotes the average of the enclosed
over the pulse's entire spatiotemporal profile; a laser beam with a Gaussian
radial and temporal profile is assumed. A complementary paper \cite%
{Ruden25Qmodn} calculates the optical cycle-average ionization rate $%
\left\langle W\right\rangle _{\gamma }$ of Ar, O$_{2}$, and N$_{2}$ vs.\
local (at a single point in spacetime) laser intensity $I$ by inverting
published ion count data \cite{Guo98}. \textquotedblleft $\left\langle
{}\right\rangle _{\gamma }$\textquotedblright\ (now with subscript $\gamma $%
) denotes the local average of the enclosed over a single optical cycle with 
$I$ identified by Keldysh parameter $\gamma $ \cite{Keldysh65} (defined
below).

Both papers are required by a third paper \cite{Ruden25QmodS} to estimate $%
I_{0}$, electron density $n$, $T$, axial current decay rate, and the
far-field microwave radiation pattern of plasma filaments formed in air for
a range of background pressures, based on measurements of electrical
conductivity (a function of both $n$ and $T$), radius, and the time integral
of axial current along the filaments' length. The filaments are formed by
geometrical convergence and subsequent Kerr focusing \cite{Sprangle14} of a
linearly polarized $800$ nm Ti:sapphire USPL pulse. Such radiation in the
microwave \cite{Englesbe18}\cite{Englesbe21} and THz \cite{Amico08} bands
has been observed \cite{Englesbe18}\cite{Englesbe21} and simulated \cite%
{Garrett21}\cite{Sprangle04} for comparison. Although the original
motivation for writing this paper and its complement \cite{Ruden25Qmodn} was
to support the third paper \cite{Ruden25QmodS}, they are each written to
stand alone due to their broader significance in supplying initial
conditions for other plasma electrodynamic simulations, and since the
results of one do not depend on the other.

In this paper, we use published electron kinetic energy spectral
measurements of a focused USPL pulse traveling through O$_{2}$ at
sufficiently low molecular density $n_{0}$ so as not to perturb the
spatiotemporal $I$ profile. The spectra are used in conjunction with a
nonadiabatic strong field approximation (SFA) of the ionization process and
a simplified model of classical rescatter, which accounts for electrons
returning to their parent ion, to estimate $T$ and $\left\langle p_{{\mathrm{%
f}z}}\right\rangle $. Modeling O$_{2}$ suffices for the atmospheric
filamentation application since N$_{2}$ only contributes 15\% of the free
electrons at $I=100$ TW/cm$^{2}$ \cite{Ruden25QmodS} (the typical magnitude
of $I$ for such filaments).\ Beyond this, and for other USPL targets, the
procedures presented may be applied to other gas species.

Our SFA model assumes that free electron energies are sufficiently high that
the Heisenberg uncertainty principle allows for a quasiclassical intracycle
temporal treatment of the ionization process, where only the most probable
tunnel path for a given instantaneous ionization time $t_{0}$ needs to be
considered. The instantaneous ionization rate $W$ (within an $I$-dependent
scaling factor), momentum vector of the electron immediately upon ionization 
$\mathbf{p}_{0}$ (referred to as \textquotedblleft residual
momentum\textquotedblright ), and the momentum's post-optical value $\mathbf{%
p}_{{\mathrm{f}}}$ are thereby found as functions of $t_{0}$. The
uncertainty principle implies that if the post-optical kinetic energy $p_{{%
\mathrm{f}}}^{2}/\left( 2m\right) $ is less than $\hbar \omega =1.55$ eV
(photon energy), where $m$ is electron mass and $\omega $ is the laser's
optical angular frequency, then $t_{0}$ cannot be specified to intracycle
precision, and less probable paths will significantly reduce the accuracy of
our SFA. Nonetheless, we will carry such cases through since a
quasiclassical solution to a quantum mechanical problem may lead to useful
insights.

A time $t$ dependent but \emph{steady state} laser pulse $\mathbf{E}$ field
amplitude envelope $\mathcal{E=E}\left( t-z/c\right) $ and its associated $%
\mathbf{B}$ field are assumed, where $c$ is the speed of light in vacuum. $%
\mathbf{E}$ and $\mathbf{B}$ are considered to be at $z=0$ unless otherwise
specified. \textquotedblleft Steady state\textquotedblright , here, means
that the waveform's axial dependence is exclusively a function of comoving
coordinate $z^{\prime }=z-ct$. $\ \mathcal{E}$ variation is assumed to be
sufficiently slow relative to $\omega $ that it may be considered constant
for intracycle operations. The laser intensity at $z=$ $0$ is $I$ $=$ $%
I\left( t\right) =\left( 1+\varepsilon ^{2}\right) c\epsilon _{0}\mathcal{E}%
^{2}/2=\left( 1+\varepsilon ^{2}\right) \omega ^{2}mU_{0}c\epsilon
_{0}/\left( e^{2}\gamma ^{2}\right) $. It is identified either by $I$
itself, $\mathcal{E}$, or the unitless Keldysh parameter \cite{Keldysh65} $%
\gamma =\omega \sqrt{2mU_{0}}/\left( e\mathcal{E}\right) $. Here, $e$ is
elemental charge, $\epsilon _{0}$ is the permittivity of free space, $U_{0}$
is ionization potential, and $\varepsilon $ is laser polarization
ellipticity as used in Eqs.~\ref{A01} of Appx.~A. SI\ units are used
throughout, except that, by convention, temperatures and energies are
converted to electron volts wherever numerical results are reported. The
nonadiabatic most probable path SFA, as formulated by Li, et al.\ \cite{Li17}
and Luo, et al. \ \cite{Luo19}, provides the basis for our model prior to
rescatter. For brevity, published references will generally be referred to
by the last name of the first author, after it has been fully cited. This
SFA's reformulation is the subject of Appx.~A, and referred to in the main
text as SFA0. Both linear ($\varepsilon =0$) and circular ($\varepsilon =1$)
polarized light cases are considered for SFA0, but the sparsity of data
limits follow-on results to $\varepsilon =0$. Preliminary results for $%
\varepsilon =1$ are included for a time when more spectral data become
available for it.

$W$ vs.\ $t_{0}$ assumed by SFA0 is, from Eq.~\ref{A09}, with $G=G_{\mathrm{c%
}}$,$\ $%
\begin{equation}
W=C_{\gamma }C_{0}\exp \left( -2G_{\mathrm{c}}/\hbar \right)   \label{02}
\end{equation}%
$G_{\mathrm{c}}$ is the imaginary part of the action of the most probable
electron tunneling path from the ground to free states with EM\ plane wave
excitation, $C_{0}$ (Eq.~\ref{A27}.1) is Li's Coulomb correction factor \cite%
{Li17}, and $C_{\gamma }$ is a $\gamma $ dependent scaling factor set to
that needed to make $W$ consistent with $\left\langle W\right\rangle
_{\gamma }$. The number after the period in equation number references here
refers to the line number of a multiline equation set that the equation
begins on. The time dependence of such\ terms is due to their dependence on $%
\gamma =\gamma \left( t\right) $ (where, recall, $t$ varies slowly). For $W=W
$ $\left( \gamma \left( t\right) ,t_{0}\right) $, and an arbitrary function $%
f=f$ $\left( \gamma \left( t\right) ,t_{0}\right) $\ of\ both $\gamma \left(
t\right) $ and intracycle ionization time $t_{0}$, we have, then,%
\begin{equation}
\begin{tabular}{l}
$\left\langle W\right\rangle _{\gamma }=\frac{\omega }{2\pi }%
\int\limits_{0}^{2\pi /\omega }Wdt_{0}$ \\ 
$\left\langle f\right\rangle _{\gamma }=\frac{\int\limits_{0}^{2\pi /\omega
}fWdt_{0}}{\int\limits_{0}^{2\pi /\omega }Wdt_{0}}=\frac{\omega }{2\pi
\left\langle W\right\rangle _{\gamma }}\int\limits_{0}^{2\pi /\omega
}fWdt_{0}$%
\end{tabular}%
\ \   \label{01}
\end{equation}%
where weak ionization is assumed throughout. $C_{\gamma }$ is not needed for
expressions of the form $\left\langle f\right\rangle _{\gamma }$ in this
paper, though, since (unlike $C_{0}$) it does not depend on $t_{0}$ and
cancels out.

The electron kinetic energy spectrum of SFA0 for $\varepsilon =0$ is used as
a fitting function to best match the spectral shape and thermalized $T$ in
cases for which spectral data are available. Two discrepancies between SFA0
and published spectra for $\varepsilon =0$ are a surge in the electron
population as kinetic energy decreases below a few eV seen in SFA0 but not
the data, and an upturn in the high energy tail (referred to as a
\textquotedblleft plateau\textquotedblright\ \cite{Becker18}) above a few
tens of eV seen in the data but not SFA0. These are interpreted to be the
result of many electrons returning to their parent ion within an optical
cycle and either recombining \cite{Talebpour96} or rescattering off of it 
\cite{Becker14}. If the latter, the electrons often achieve a higher
post-optical energy than would otherwise be obtained. We address the absence
of the surge in Sec.~II with a model referred to as SFA1 by replacing the
unobserved surge with a ceiling limiting the spectral amplitude of electrons
below a critical energy that fits the empirical spectra better.
Energy-boosted electrons that help deplete this surge form the plateau. We
then use a phenomenological energy multiplier for SFA1 spectra in Sec.~II as
a second fitting parameter to match the higher $T$ that results for model
SFA2. Since SFA1 and SFA2 entail fits to data from optical pulses that only 
\emph{peak} in intensity at specified values of $\gamma =\gamma _{0}$ (or,
equivalently, $\mathcal{E=E}_{0}$ and $I\mathcal{=}I_{0}$), their results
are interpreted to be functions representative of a full pulse's $\gamma
_{0} $, instead of SFA0's uniform constant $\gamma $.

We go beyond Li \cite{Li17} and Luo \cite{Luo19} in Appx.~A by solving for
residual and post-optical $p_{0z}$ and $p_{{\mathrm{f}z}}$ (Eqs.~\ref{A24}),
respectively, as second order terms. They result from our first order
estimate of the momentum vector transverse to the $z$-axis $\mathbf{p}_{r}$
crossing the optical magnetic field $\mathbf{B}$ (the optical magnetic
Lorentz force) during and after tunneling, respectively. This is neglected
in their work since $p_{z}\ll p_{r}$, so has little effect on the kinetic
energy spectrum (their primary interest). It is needed for plasma
electrodynamic\ applications, though, since it (along with $\left\langle
W\right\rangle _{\gamma }$ \cite{Ruden25Qmodn}) is needed to determine
post-optical current density $\mathbf{J}$. $p_{r}$, here, is the (radial)
component of momentum transverse to the $z$ axis ($p_{r}=p_{x}$ for $%
\varepsilon =0$).

SFA2 assumes a semiclassical three-step process (quasi-classical ionization,
classical dynamics as a free electron, then classical rescatter to the
detector) that reportedly \textquotedblleft embodies the generally accepted
physical picture of the HATI [High-order Above-Threshold Ionization]
process\textquotedblright\ \cite{Becker18}. As has been suggested \cite%
{Becker14}, a simplified model of the classical kinematics of rescatter is
used in Sec.~III to infer the electron mean kinetic energy angular
distribution to first order. $\left\langle p_{{\mathrm{f}z}}\right\rangle $
is much too small for spectral measurements to resolve directly, though.
Instead, we calculate it by averaging it as a second order effect as before,
but with contributions before and after rescatter. A more complete HATI
treatment entails solving for the full SFA scattering matrix (not just the
most probable path) with rescatter included \cite{Busuladzic08a}. This is
computationally demanding for diatomic molecules such as O$_{2}$ \cite%
{Busuladzic08b} since the effect of the molecular potential at all rotation
angles must be averaged over. We take instead our semi-empirical approach
since our purpose is to make the best use of available spectral data for
post-optical plasma physics simulations that accept $T$ and $\mathbf{J}$ as
initial conditions vs.\ advancing scattering theory. No reports are found of
HATI results solving for the resultant thermalized $T$ or $\left\langle p_{{%
\mathrm{f}z}}\right\rangle $ by spectral averaging for comparison. $p_{{%
\mathrm{f}z}}$ is usually neglected entirely. A degree of validation for our 
$\left\langle p_{{\mathrm{f}z}}\right\rangle $ result is provided by
consistency of the inferred far-field microwave emission pattern of a
filament with actual measurements of such. \cite{Ruden25QmodS}

\section{Electron kinetic energy spectra and thermalized temperature}

Plots of the total momentum transfer cross sections for electron scattering
off neutral N$_{2}$ \cite{Kawaguchi21} and O$_{2}$ \cite{Kawaguchi25}, and
their elastic scattering contributions show that the latter dominates
sufficiently that there is sufficient of time for the electron energy
spectrum to thermalize prior to significant energy depletion in the $0.1$ to 
$10$ eV range of interest. Meanwhile, effective momentum transfer collision
frequency for a thermalized (Maxwellian) electron energy distribution for an
air mixture at 1 atm (molecular density $n_{\text{a}}=2.6868\times 10^{25%
\text{ }}$m$^{-3}=$ Loschmidt's number \cite{NRL}) is \cite{Pusateri15} $%
6.\,\allowbreak 2\times 10^{12}$ s$^{-1}$. This is sufficiently slow
relative to the USPL pulse that the post-optical phase is temporally
well-separated from the thermalized phase.

We see from Fig.~A3a that the radial component of post-optical momentum $p_{{%
\mathrm{f}r}}$ and, therefore, post-optical electron kinetic energy $U$
(where subscript \textquotedblleft f\textquotedblright\ is suppressed) are
monotonically increasing functions of $t_{0}$ for $\varepsilon =0$ over the
first $t_{0}$ quarter-cycle, and that this represents the entire spectrum. $%
\left\langle f\left( U\right) \right\rangle _{\gamma }$ for any function $%
f\left( U\right) $ of $U$ over the first quarter-cycle of SFA0, therefore,
is the integral of $f\left( U\right) S_{0}\left( U\right) $ over $U$\ from $%
U=0$ to $\infty $, were $S_{0}\left( U\right) $ is the $U$ spectrum for
SFA0. This allows us to identify $S_{0}\left( U\right) $ for $\varepsilon =0$
in the following by changing\ the integration variable from\ $t_{0}$ to $U$
in the expression for $\left\langle U\right\rangle _{\gamma }$, based on Eq.~%
\ref{01}.2,%
\begin{equation}
\begin{tabular}{l}
$\varepsilon =1$:\ \ $\frac{3}{2}k_{\mathrm{B}}T_{0}=U=\frac{p_{{\mathrm{f}r}%
}^{2}}{2m}$ \\ 
$S_{0}\left( U\right) =\delta \left( U-p_{{\mathrm{f}r}}^{2}/\left(
2m\right) \right) $ \\ 
$\varepsilon =0$:\ \ $\frac{3}{2}k_{\mathrm{B}}T_{0}=\left\langle
U\right\rangle _{\gamma }=\frac{\int\limits_{0}^{\pi /2}UWd\tau _{0}}{%
\int\limits_{0}^{\pi /2}Wd\tau _{0}}$ \\ 
$=\int\limits_{0}^{+\infty }US_{0}\left( U\right) dU$ \ \ \ \ \ \ \ \ \ $%
\tau _{\left[ n\right] }=\omega t_{\left[ n\right] }$ \\ 
$S_{0}\left( U\right) =\frac{W\left( \frac{dU}{d\tau _{0}}\right) ^{-1}}{%
\int\limits_{0}^{\pi /2}Wd\tau _{0}}$ where \ $\tau _{0}\rightarrow \tau
_{0}\left( U\right) $%
\end{tabular}%
\ \ \ \   \label{03}
\end{equation}%
$k_{\mathrm{B}}$ is Boltzmann's constant and $T_{0}$ is the temperature
implied by SFA0 upon thermalization of $S_{0}\left( U\right) $.

$U=p_{{\mathrm{f}r}}^{2}\ /\left( 2m\right) $ vs.\ $\tau_{0}$ in the above
is from Eqs.~\ref{A15} for both polarities. The contribution to $U$ from $p_{%
{\mathrm{f}z}}^{2}/\left( 2m\right) $ is negligible, so is neglected for our 
$T_{0}$ calculation. $U$ for $\varepsilon=1$ (circular polarization) has no $%
\tau_{0}$ dependence, so its $S_{0}\left( U\right) $ is represented by a
Dirac delta function in Eq.~\ref{03}.2. For $\varepsilon=0$, $\tau
_{0}\rightarrow\tau_{0}\left( U\right) $ in Eq.~\ref{03}.5 specifies that $%
\tau_{0}$ dependences of the numerator be expressed as function $\tau
_{0}\left( U\right) $ of the new integration variable $U$ \emph{after} $%
dU/d\tau_{0}$ is solved for. This $\tau_{0}\left( U\right) $ is found by
substituting the expression for $p_{{\mathrm{f}r}}$ in Eq.~\ref{A15}.3 into $%
U=p_{{\mathrm{f}r}}^{2}/\left( 2m\right) $ and solving for $\tau_{0}$. The
results, for use in Eqs.~\ref{03}, are determined from the following,%
\begin{equation}
\begin{tabular}{l}
$\varepsilon=1$:\ \ $U=\frac{U_{0}}{\gamma^{2}}\left( \cosh\tau_{\mathrm{ic}%
}-\sqrt{\sinh^{2}\tau_{\mathrm{ic}}-\gamma^{2}}\right) ^{2}$ \\ 
$\varepsilon=0$:\ \ $W=C_{\gamma}C_{0}\exp\left( -\frac{2G_{\mathrm{c}}}{%
\hbar}\right) $ \\ 
$U=\frac{U_{0}\sin^{2}\tau_{0}}{\gamma^{2}}\left( 1+\frac{\gamma^{2}}{%
\cos^{2}\tau_{0}}\right) $ \\ 
$\frac{dU}{d\tau_{0}}=\frac{2U_{0}\sin\tau_{0}\left(
\gamma^{2}+\cos^{4}\tau_{0}\right) }{\gamma^{2}\cos^{3}\tau_{0}}$ \\ 
$\cos^{2}\left( \tau_{0}\left( U\right) \right) =\frac{1}{2}-\frac {%
\gamma^{2}}{2}-\frac{\gamma^{2}U}{2U_{0}}$ \\ 
$+\frac{1}{2}\sqrt{\frac{\gamma^{4}U^{2}}{U_{0}^{2}}+\frac{2\gamma^{2}\left(
\gamma^{2}-1\right) U}{U_{0}}+\gamma^{4}+2\gamma^{2}+1}$%
\end{tabular}
\label{04}
\end{equation}
where Eqs.~\ref{A27} and Eq.~\ref{A14}.1 are used for $C_{0}$ and $G_{%
\mathrm{c}}$, respectively, for $\varepsilon=0$, and $\tau _{\mathrm{ic}}$
is determined numerically from the minimum of Eq.~\ref{A13}.1 for $%
\varepsilon=1 $.

Several published spectra are used to adjust SFA0 to better fit the data.
Figure~1 plots a representative few, but all found are represented in the
thermalized $T$ plots of Fig.~2. Table 1 columns list, respectively, the
value of $\gamma _{0}$, the directions the electrons are averaged over, the
pulse width of the laser (fs), and the reference. Re directions,
\textquotedblleft $x$\textquotedblright\ in Table 1 refers to the $x$%
-direction only, \textquotedblleft $2\pi $\textquotedblright\ to all $2\pi $
rad in the $x$-$y$ plane, and \textquotedblleft $4\pi $\textquotedblright\
to all $4\pi $ sr. The $\gamma _{0}$ values with superscript
\textquotedblleft $\ast $\textquotedblright\ are based on spectra that have
not converged to zero at the largest $U$ plotted, resulting in a $T$
estimate that is lower than actual. Those with superscript \textquotedblleft 
$\dag $\textquotedblright\ are for $\varepsilon =1$, with the rest being for 
$\varepsilon =0$. The spectral data is too sparse to assess trends for $%
\varepsilon =1$.

\textbf{Table 1 }Empirical spectral parameters%
\begin{equation*}
\begin{tabular}{llll|llll}
$\gamma_{0}$ & dir & fs & ref & $\gamma_{0}$ & dir & fs & ref \\ \hline
$0.47$ & $x$ & $100$ & \cite{Okunishi07} & $0.63$ & $2\pi$ & $100$ & \cite%
{Okunishi08} \\ 
$0.58$ & $x$ & $100$ & \cite{Okunishi07} & $0.77$ & $2\pi$ & $100$ & \cite%
{Okunishi08} \\ 
$0.74$ & $x$ & $100$ & \cite{Okunishi07} & $0.96$ & $2\pi$ & $100$ & \cite%
{Okunishi08} \\ 
$0.90$ & $x$ & $100$ & \cite{Okunishi07} & $1.24^{\ast}$ & $2\pi$ & $100$ & 
\cite{Okunishi08} \\ 
$1.06$ & $x$ & $100$ & \cite{Okunishi07} & $0.71^{\ast}$ & $4\pi$ & $24$ & 
\cite{Wu13} \\ 
$1.40$ & $x$ & $100$ & \cite{Okunishi07} & $0.82$ & $4\pi$ & $40$ & \cite%
{Yu20} \\ 
$2.00^{\ast}$ & $x$ & $45$ & \cite{Kloda10} & $1.30$ & $4\pi$ & $25$ & \cite%
{Deng11} \\ 
$2.34\allowbreak$ & $x$ & $45$ & \cite{Kloda10} & $1.06^{\dag}$ & $x$ & $100$
& \cite{Okunishi07} \\ 
$3.82$ & $x$ & $45$ & \cite{Kloda10} & $1.30^{\dag}$ & $4\pi$ & $25$ & \cite%
{Deng11}%
\end{tabular}
\ \ \ \ \ 
\end{equation*}

\begin{figure}[H]\centering\includegraphics[width=3.375in]{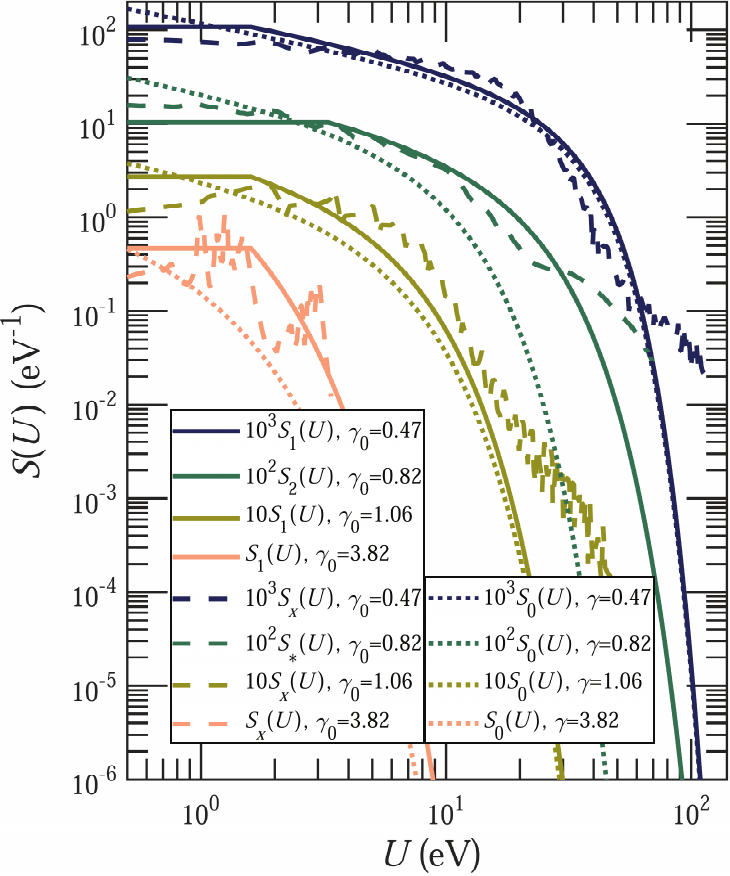}\end{figure}    
\textbf{Fig.~1 \ } Theoretical electron kinetic energy spectra $S_{0}\left(
U\right) $ for our baseline model SFA0 for O$_{2}$ exposed to an 800 nm
linearly polarized ($\varepsilon=0$) USPL (dotted lines) and SFA1's spectrum 
$S_{1}\left( U\right) $, where a ceiling of $S_{0}\left( U_{\mathrm{c}%
}\right) $ has been placed on $S_{0}\left( U\right) $ (solid lines), are
overlaid onto measured spectra $S_{x}\left( U\right) $ (dashed lines)
recorded for electrons emitted in the ($\mathbf{E}$ field) $x$-direction,
except for $\gamma_{0}=0.82$. The solid line\ for $\gamma_{0}=0.82$ is
SFA2's $S_{2}\left( U\right) =\zeta S_{1}\left( U/\zeta\right) $. It is
overlaid onto $S_{\ast}\left( U\right) $ measured for electrons averaged
over $4\pi$ sr, to which $S_{2}\left( U\right) $ better fits. The color
scheme here and elsewhere is that of batlow10 \cite{Crameri20} to enable
gray scale interpretation. The plots of different colors (vertically
separated by prefactors) are for different values of $\gamma_{0}$ (minimum $%
\gamma$).

\bigskip

Two modifications to $S_{0}\left( U\right) $ for $\varepsilon =0$ are used
to approximate the empirical spectra and their thermalized temperatures.
They are symbolized by incrementing their \emph{numerical} subscript. For
SFA1, $S_{1}\left( U\right) $ and $T_{1}$ result from a ceiling placed on $%
S_{0}\left( U\right) $. For SFA2, $S_{2}\left( U\right) $ and $T_{2}$ result
from an energy rescaling of $S_{1}\left( U\right) $. Data to which they are
fit, meanwhile, are represented by \emph{non-numerical} subscripts: $x$, $+$%
, and $\ast $, representing published spectra recorded from electrons
emitted in the $x$-direction, averaged over all $2\pi $ radians in the $x$-$%
y $ plane, and over all $4\pi $ sr, respectively.

Figure~1 plots $S_{0}\left( U\right) $ vs.\ $U$ for O$_{2}$ ($U_{0}=12.063$
eV \cite{Samson66}) exposed to $800$~nm light ($\hbar\omega=1.550$ eV) for
values of $\gamma_{0}$ for which there are experimental spectra to compare
(referenced in Table~1) for $\varepsilon=0$. Figure~1 also plots examples of
the adjusted theoretical spectra $S_{1}\left( U\right) $ and $S_{2}\left(
U\right) $ overlaid on their corresponding empirical counterparts with the
same $\gamma_{0}$. The spectra are all specific to O$_{2}$ exposed to $800$
nm\ light, for which $N_{\mathrm{q}}=7.78$ and $\kappa=0.94$ from Eq.~\ref%
{A27}.3.

\begin{figure}[H]\centering\includegraphics[width=3.375in]{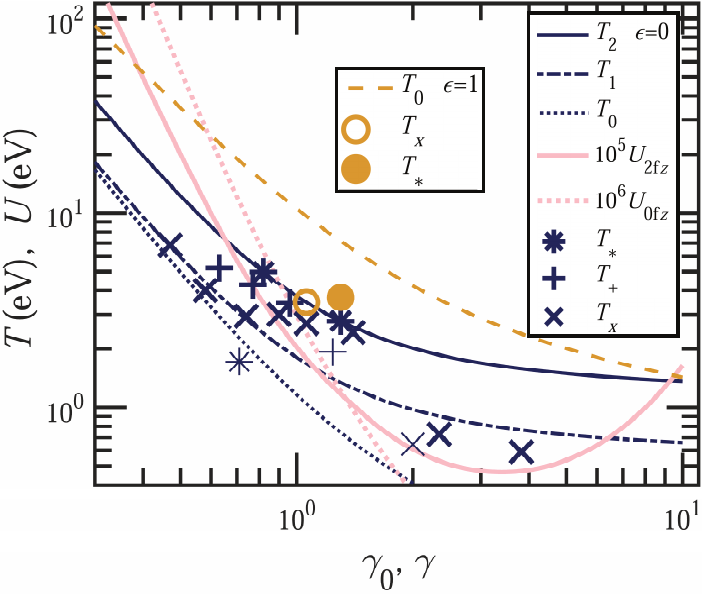}\end{figure}   
\textbf{Fig.~2} $T$ vs.\ $\gamma_{0}$ (vs.\ $\gamma$ for $T_{0}$ and $U_{0{%
\mathrm{f}z}}$) for O$_{2}$ illuminated by $800$ nm after thermalization of
the initial electron kinetic energy spectrum showing the temperatures
resulting from thermalizing $S_{0}\left( U\right) $, $S_{1}\left( U\right) $%
, and $S_{2}\left( U\right) $, defined as $T_{0}$, $T_{1}$, and $T_{2}$,
respectively, are plotted for $\varepsilon=0$. Experimental results are
overlaid for electrons emitted in the ($\mathbf{E}$ field) $x$-direction ($%
T_{x}$), averaged over all $2\pi$ radians in the $x$-$y$ plane ($T_{+}$),
and over all $4\pi$ sr ($T_{\ast}$). $T_{0}$ and $S_{0}\left( U\right) $ for 
$\varepsilon=1$, and the kinetic energies $U_{2{\mathrm{f}z}}$ and $U_{0{%
\mathrm{f}z}}$ of an electron with mean axial momentum for SFA2 (with
rescatter) and SFA0 (without rescatter), are also plotted. Note the large
and different scaling factors needed to put $U_{2{\mathrm{f}z}}$ and $U_{0{%
\mathrm{f}z}}$ on-scale.

\bigskip

Figure~2 plots $T$ vs.\ $\gamma _{0}$ for O$_{2}$ exposed to $800$ nm after
thermalization of the electron kinetic energy spectrum. The markers for $T$
representing incomplete spectra (superscript $\ast $\ in Table 1) are
plotted for reference, but have thinner lines and are not used for the fits.
A significant difference between $S_{1}\left( U\right) $ and $S_{\ast
}\left( U\right) $ seen in Fig.~1 is that the latter displays what is
attributed to a rescatter \textquotedblleft plateau\textquotedblright\ \cite%
{Becker18} at higher $U$ not seen in $S_{1}\left( U\right) $. The
correlation is better between $S_{1}\left( U\right) $ and $S_{x}\left(
U\right) $, though, for $0.47\leq \gamma _{0}\leq 0.74$, with good agreement
between $T_{1}$ and $T_{x}$ in this range shown in the plot. However, $T_{x}$
rises above $T_{1}$ over the interval $0.74\leq $ $\gamma _{0}\leq 1.40$,
whereupon it approximates $T_{\ast }$ at $\gamma _{0}=1.40$. The $\gamma
_{0}=1.06$ plots in Fig.~1 compare $S_{1}\left( U\right) $ to $S_{x}\left(
U\right) $ midway within this transition interval to illustrate this
departure.

Re $S_{1}\left( U\right) $, there is a surge in the electron population for $%
\varepsilon=0$ in $S_{0}\left( U\right) $ for low $U$. This is due to the
highest ionization rate being at the peak $\mathbf{E}$ field magnitude, just
when $p_{{\mathrm{f}r}}$ (and therefore $U$) approaches zero in SFA0 (Eq.~%
\ref{A15}.3 and Fig.~A3a). However, this surge is not seen in the
experimental spectra $S_{x}\left( U\right) $ (collected in the $\mathbf{E}$
field direction), $S_{+}\left( U\right) $ (averaged over $2\pi$ rad in the $%
x $-$y$ plane), or $S_{\ast}\left( U\right) $ (averaged over $4\pi$ sr).
Correlation with $S_{x}\left( U\right) $ is improved for $\gamma_{0}\leq0.74$
by placing a ceiling on $S_{0}\left( U\right) $ by setting its value for $%
U<U_{\mathrm{c}}$ equal to $S_{0}\left( U_{\mathrm{c}}\right) $, and then
renormalizing it so that it still integrates to unity. We define this as $%
S_{1}\left( U\right) $. The value chosen of $U_{\mathrm{c}}=1.60$ eV is
discussed below.

Re $S_{2}\left( U\right) $, a second fitting parameter $\zeta=2.067$ is used
to better fit our model to $T_{\ast}$. For this, $S_{1}\left( U\right) $ is
replaced by $S_{2}\left( U\right) =S_{1}\left( U/\zeta\right) /\zeta$. The
values of $\zeta=2.067$ and the $S_{1}\left( U\right) $ ceiling energy $U_{%
\mathrm{c}}=1.60$ eV are chosen to match $T_{2}$ to $T_{\ast}$ at $%
\gamma_{0}=0.82$ and $\gamma_{0}=1.30$, where good data is available. $T_{2}$
for $0.82\leq$ $\gamma_{0}\leq1.30$ may therefore be considered the most
accurate range of $T_{2}$ since it is simply an interpolation between data
points. Additional spectra averaged over $4\pi$ sr would clarify $T_{2}$'s
accuracy for $\gamma_{0}\leq0.82$. Short of this, the good fit of $T_{1}$ to 
$T_{x}$ for $0.47\leq$ $\gamma_{0}\leq0.74$, and the trend displayed\ by $%
T_{+}$ consistent with being intermediate between $T_{x}$ and $T_{\ast}$,
suggests that $T_{2}$ may be a useful approximation down to $\gamma_{0}=0.47$%
.

The error resulting from extrapolating $T_{2}$ to $1.4<\gamma_{0}<2.34$ is
unknown. For $\gamma_{0}\geq2.34$, though, we see in Fig.~1 that $U$ for
most electrons falls below $U_{\mathrm{c}}$, so SFA2 is likely of little
value. This is the multiphoton ionization regime, where ionization depends
on the details of electronic structure not well-represented by any SFA.

\section{Mean post-optical axial momentum $\left\langle p_{2{\mathrm{f}z}%
}\right\rangle $ based on the kinematics of rescatter}

This section is limited to $\varepsilon =0$ due to our inability to model $%
\varepsilon =1$. We assume here that SFA0 is valid in the absence of the
recombination of electrons with and elastic rescatter off of their parent
ion, and that the two fitting parameters resulting in $S_{2}\left( U\right) $
correct for these. The electron's mean post-optical axial moment $%
\left\langle p_{2{\mathrm{f}z}}\right\rangle $ (with subscript $2$ referring
to SFA2 specifically now) is inferred from the classical kinematics of
rescatter subsequent to an initial ionization described by SFA0, but with a
post-optical angular distribution consistent with the empirical data
presented in the previous section. Additional subscripts $\phi $ and $y$ are
introduced, representing properties of electrons emitted at angle $\phi $
relative to the $x$-axis \emph{after} pulse passage, and in the $y$%
-direction, respectively. When not used to identify SFA2 properties,
subscripts $1$ and $2$ in this section refer to properties immediately
before and after rescatter, respectively. If, however, subscript $2$ is
followed by an \textquotedblleft f\textquotedblright , \ that refers to a
post-optical property \emph{affected} by recombination/rescatter.

If post-optical emission were strictly in the $x$-direction, then $%
T_{x}=T_{\ast }$, and the spatiotemporal average of $p_{{\mathrm{f}z}}$ for
a full optical pulse is, from Eqs.~\ref{A24}, 
\begin{equation}
\left\langle p_{{\mathrm{f}z}}\right\rangle =\frac{\left\langle p_{{\mathrm{f%
}x}}^{2}\right\rangle }{2mc}=\frac{3k_{\mathrm{B}}T_{x}}{2c}  \label{05}
\end{equation}%
We infer from Fig.~2 and published experimental plots \cite{Okunishi08},
though, that there is a significant rate of post-optical electrons emitted
per unit solid angle at large angles $\phi $, at least for $\gamma _{0}\sim
1 $, and that the $U$ of such electrons rises with $\phi $. This has been
interpreted to be due to electrons rescattering off of their parent ion \cite%
{Becker14} in the oscillating laser $\mathbf{E}$ field, and subsequently
receiving a higher post-optical $U$. As such, it is assumed to be symmetric
about the $x$-axis to first order. $p_{y}$, therefore, represents the first
order component of momentum in \emph{any} direction orthogonal to the $x$%
-axis (not just $y$).

Rescattered electrons are assumed to be accelerated by the optical $\mathbf{E%
}$ exclusively between release and rescatter (Coulomb interaction of the
parent is neglected). Confining ourselves now to real classical mechanics,
we define $x_{r}\left( t\right) $ as the real first order $x$-coordinate of
the electron after it is freed at $t=t_{0}$. Given $x_{r}\left( t_{0}\right)
=-r_{00}\left( 0\right) $ from Eq.~\ref{A26}.8 and $dx_{r}\left( t\right)
/dt=p_{r}/m$ from Eq.~\ref{A04}, integrating the latter from $t_{0}$ to $t$
results in,%
\begin{equation}
x_{r}\left( t\right) =\left( t-t_{0}\right) \frac{p_{{\mathrm{f}x}}}{m}%
-r_{00}\left( 0\right) +\frac{e}{m}\int\limits_{t_{0}}^{t}\mathbf{A}%
_{0}\left( t^{\prime }\right) dt^{\prime }  \label{06}
\end{equation}%
where $\mathbf{A}_{0}\left( t\right) =\mathbf{A}$ of Eq.~\ref{A01}.1 with $%
\varepsilon =z=0$ and $p_{{\mathrm{f}r}}=p_{{\mathrm{f}x}}$. The electron
returns to the nucleus at $x=0$ at time $t_{1}$, implying $x_{r}\left(
t_{1}\right) =0$. Substituting expressions for $r_{00}\left( 0\right) $, $p_{%
{\mathrm{f}x}}$, and $\mathbf{A}_{0}\left( t\right) $ from Eq.~\ref{A26}.8,
Eq.~\ref{A15}.3, and Eq.~\ref{A01}.1, respectively, and changing variables
to $\gamma $ from Eq.~\ref{A13}.8 and $\tau _{\left[ n\right] }=\omega t_{%
\left[ n\right] }$, Eq.~\ref{06} normalizes to,%
\begin{equation}
0=\left( \left( \tau _{1}-\tau _{0}\right) \sin \tau _{0}-\cos \tau
_{0}\right) \sqrt{1+\frac{\gamma ^{2}}{\cos ^{2}\tau _{0}}}+\cos \tau _{1}
\label{07}
\end{equation}

Newton's method for finding the zero of a function is used (with $\tau
_{0}=0 $ for the first iteration) to find the value of $\tau _{0}$ for which
the r.h.s.\ of Eq.~\ref{07} equals zero for $0.1\leq \gamma \leq 10$ and $%
0<\tau _{1}\leq 6.6$. $\tau _{0}$ solutions are rejected for any given $%
\gamma $\ for all values less than the greatest value $\tau _{0,\min }$ for
which iterations do \emph{not} converge\ (implying the electron does not
return to its parent), \emph{or} if $\tau _{0}\geq \tau _{0,\max }=1.4$
(beyond which the ionization rate is negligible). There are no physically
meaningful solutions (with $\tau _{1}>\tau _{0}$) for the second $\tau _{0}$
quarter-cycle ($\pi /2\leq \tau _{0}<\pi $). This is a feature shared by the
so-called \textquotedblleft Simple Man Model\textquotedblright\ (SMM)\ \cite%
{Becker14}, for which $x_{r}\left( t_{0}\right) =p_{0x}=0$. Upon integrating
the electron equation of motion with these initial conditions, one finds the
relationship between $\tau _{1}$ and $\tau _{0}$ for the SMM is Eq.~\ref{07}
with $\gamma =0$. Although solutions exist for both models in the second
half-cycle, the most probable path that we limit ourselves to is already
covered by the kinematics of the first half-cycle, but in the opposite
direction. $\tau _{1}$ vs.\ $\tau _{0}$ is plotted in Fig.~3a for a range of 
$\gamma $ values.

\begin{figure}[H]\centering\includegraphics[width=3.375in]{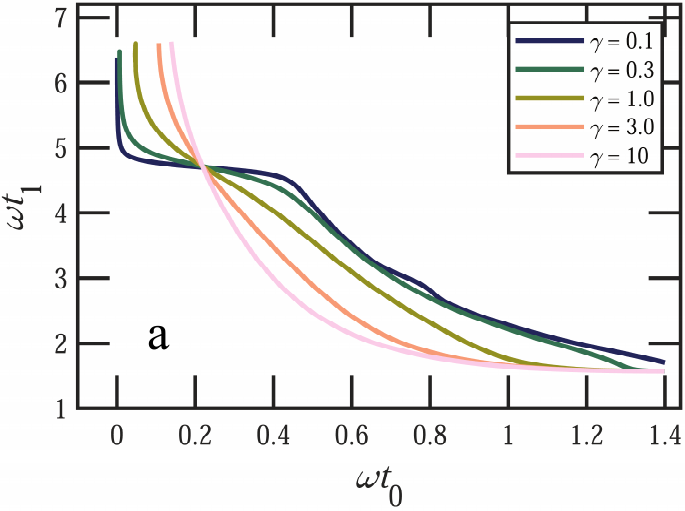}\end{figure}     
\begin{figure}[H]\centering\includegraphics[width=3.375in]{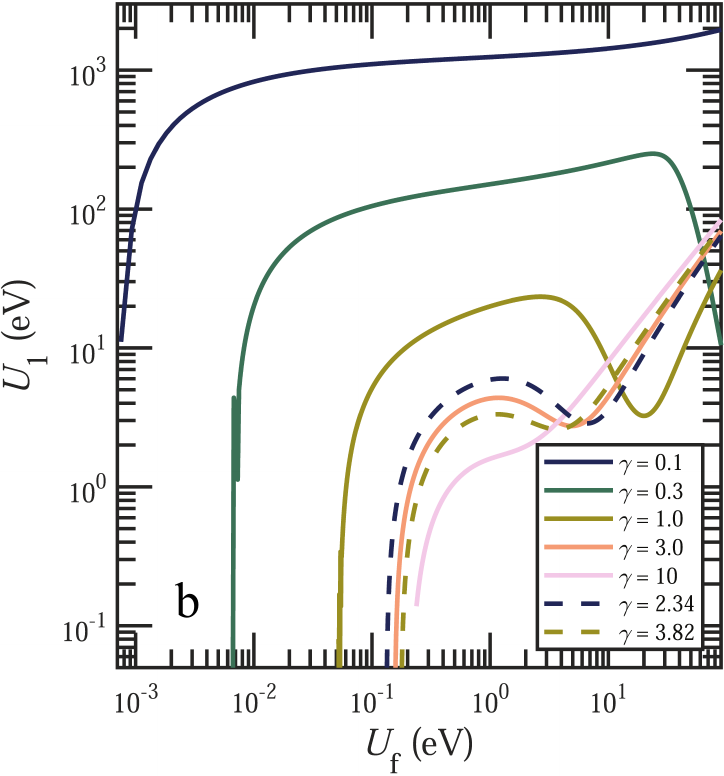}\end{figure}     
\textbf{Fig.~3} SFA0's electron-parent ion re-encounter time $t_{1}$ vs.\
ionization time $t_{0}$ for $\varepsilon =0$ (a) and electron kinetic energy 
$U_{1}=p_{1x}^{2}/\left( 2m\right) $ at $t=t_{1}$ vs.\ its post-optical
kinetic energy in the absence of parent interaction $U_{{\mathrm{f}}}=p_{{%
\mathrm{f}x}}^{2}/\left( 2m\right) $ (b). Two special $\gamma $ cases
(dashed lines) are included for which K{\l }oda, et al.\ \cite{Kloda10} has
data relevant to the value of $U_{{\mathrm{f}}}=U_{{\mathrm{f,min}}}$, below
which the electron theoretically does not return to its parent (indicated by
the sudden drop-off in $U_{1}$). The $\tau _{1}$ vs.\ $\tau _{0}$ curves ($%
\tau =\omega t$) all cross at $\tau _{0}=0.219$ (when $\tau _{1}=3\pi /2$
and $\mathbf{E=0}$) due to $\cos \tau _{1}$ and $\left( \cos \tau
_{0}-\left( \tau _{1}-\tau _{0}\right) \sin \tau _{0}\allowbreak \right)
\allowbreak $ both being zero then.

\bigskip

$p_{x}$ at $\tau =\tau _{1}$ for $\varepsilon =0$ at $z=0$ is, from Eq.~\ref%
{A04}, Eq.~\ref{A15}.3, Eq.~\ref{A01}.1, and Eq.~\ref{A13}.8,%
\begin{equation}
p_{1x}=\frac{\sqrt{2mU_{0}}}{\gamma }\left( \sin \tau _{0}\sqrt{1+\frac{%
\gamma ^{2}}{\cos ^{2}\tau _{0}}}-\sin \tau _{1}\right)  \label{08}
\end{equation}%
Kinetic energy $U_{1}=p_{1x}^{2}/\left( 2m\right) $ upon electron return,
and its post-optical value $U_{{\mathrm{f}}}=p_{{\mathrm{f}x}}^{2}/\left(
2m\right) $ \emph{in the absence of any interaction with the parent ion}
(from Eq.~\ref{A18}.3) are then both calculated as functions of $\tau _{0}$,
and parametrically plotted\ against each other in Fig.~3b. Results for two $%
\gamma $ values of special interest are added, as discussed below. $U_{{%
\mathrm{f}}}$ in this section (where subscript \textquotedblleft
f\textquotedblright\ is no longer suppressed) refers specifically to the
value implied by SFA0.

The observed low $U$ spectral suppression likely results, at least in part,
from a decrease in $U_{1}$ (implying a higher probability of recombination)
as $U_{{\mathrm{f}}}$ decreases, as seen in Fig.~3b. Note, though, that
there is a value of $U_{{\mathrm{f}}}=U_{{\mathrm{f,}\min }}$ below which
the electron does \emph{not} return to its parent (even in the first
quarter-cycle) and, therefore, avoids recombination in this model. $U_{{%
\mathrm{f,}\min }}$ is too small to make a significant difference in $T$ for
the empirical spectra, with the exception of the $\gamma _{0}=3.82$ ($%
I_{0}=6.9$ TW/cm$^{2}$) and $\gamma _{0}=2.34$ ($I_{0}=18.4$ TW/cm$^{2}$)
spectra of K{\l }oda, et al.\ \cite{Kloda10}. Though off-scale in our
Fig.~1, the $\gamma _{0}=3.82$ spectral amplitude rises an order of
magnitude below $U=0.2$ eV in K{\l }oda's Fig.~2 as $U$ decreases (resulting
in its low $T_{x}$ value in our own Fig.~2). This $U$ approximates $U_{{%
\mathrm{f,}\min }}$ for $\gamma =3.82$ in our Fig.~3b. This very low $U$
surge is less apparent in K{\l }oda's $\gamma _{0}=2.34$ spectral plot
since, as K{\l }oda explains, it is scaled to emphasize the higher $U$
resonant peaks. The surge only extends about half as far as $\gamma =3.82$
does in terms of $U_{{\mathrm{f}}}$, consistent with $U_{{\mathrm{f,}\min }}$
for $\gamma _{0}=2.34$. Note, though, that these $\gamma $ values are well
within the multiphoton regime for which our SFA is ill-suited. Nonetheless,
the suggestion that the observed surge in\ very low $U$ electrons\ is the
result of such electrons not returning to their parent to recombine is
worthy of further investigation.

If, instead of recombining, the electron at $\tau =\tau _{1}$ elastically
rescatters to angle $\phi _{2}$, then $p_{x}$ is reinitiallized to $%
p_{2x}=p_{1x}\cos \phi _{2}$ pursuant to further acceleration. Meanwhile,
the rescattered electron momentum normal to the $x$-axis $p_{2y}=p_{1x}\sin
\phi _{2}$ stays constant to first order thereafter. The time integral of
the first order term of Eq.~\ref{A03}.2 from these reinitiallized conditions
to infinity (when $\mathbf{A}=\mathbf{0}$) implies the post-optical momentum
components at $z=0$ are revised to, 
\begin{equation}
\begin{tabular}{l}
$p_{2{\mathrm{f}x}}=\frac{\sqrt{2mU_{0}}\sin \tau _{1}}{\gamma }+p_{1x}\cos
\phi _{2}$ \\ 
$p_{2{\mathrm{f}y}}=p_{1x}\sin \phi _{2}$ \\ 
$\tan \phi =\frac{p_{2{\mathrm{f}y}}}{p_{2{\mathrm{f}x}}}=\frac{\gamma
p_{1x}\sin \phi _{2}}{\sqrt{2mU_{0}}\sin \tau _{1}+\gamma p_{1x}\cos \phi
_{2}}$%
\end{tabular}%
\ \   \label{09}
\end{equation}%
where we have used Eq.~\ref{A01}.1 and Eq.~\ref{A13}.8.

Equation \ref{07}, Eq.~\ref{08}, and Eqs.~\ref{09} provide the kinematic
basis for the high $U$ plateau, and a second proposed mechanism by which the
low $U$ spectral surge of SFA0 is reduced. For example, at the $\tau
_{0}=0.219$, $\tau _{1}=3\pi /2$ crossing point (common to all $\gamma $),
the post-optical energy $U_{2{\mathrm{f}}}=\left( p_{2{\mathrm{f}x}}^{2}+p_{2%
{\mathrm{f}y}}^{2}\right) /\left( 2m\right) $ for $\gamma =0.82$ is $%
\allowbreak \allowbreak 24.5$ eV for $\phi _{2}=\pi /3$ (forward scatter,
but at a significant angle), and $93.6$ eV for $\phi _{2}=\pi $ (full
backscatter). These energies span the rescatter plateau of the $\gamma
_{0}=0.82$ data in Fig.~1. By contrast, $U_{{\mathrm{f}}}$ is only $%
1.45\allowbreak $ eV (in the absence of rescatter with $\phi _{2}=0$). That
is, a great $U$ boost is imparted to the large population of otherwise low $%
U=U_{{\mathrm{f}}}$ electrons released near the peak of $\mathbf{E}$ (such
as when $\tau _{0}=0.219$) if rescattered at significant angles. That said,
the plateau $U$ range (Fig. 1) is orders of magnitude higher that the mean $%
U $ that determines $T$ (Fig. 2). Average properties such as $T$ and $%
\left\langle p_{{\mathrm{f}z}}\right\rangle $, then, may be adequately
determined from the Low-Energy Structure (LES) \cite{Becker14} probed by
forward scattering alone ($\phi _{2}\leq \pi /2$).

Calculating $p_{2{\mathrm{f}z}}$ is a two-step process in the case of
rescatter. Integration of Eq.~\ref{A20} with $r=x$ along the trajectory
prior to rescatter (from $\tau =\tau _{0}$ to $\tau =\tau _{1}$), and then
from rescatter to optical pulse passage ($\tau =\tau _{1}$ to $\tau =\infty $%
) give us $p_{1z}$ and $p_{2{\mathrm{f}z}}$, respectively, where,%
\begin{equation}
\begin{tabular}{l}
$p_{1z}-p_{0z}=\frac{p_{1x}^{2}}{2mc}-\frac{p_{0x}^{2}}{2mc}$ \\ 
$p_{2{\mathrm{f}z}}-p_{1z}=\frac{p_{2{\mathrm{f}x}}^{2}}{2mc}-\frac{%
p_{2x}^{2}}{2mc}$ \\ 
$p_{2x}=p_{1x}\cos \phi _{2}$%
\end{tabular}%
\ \   \label{10}
\end{equation}%
Note for line 2 that Eq.~\ref{A20} (with $r=x)$ remains valid after
rescattering since $p_{y}$ plays no role in the optical magnetic Lorentz
force for $\varepsilon =0$. From these, and the solution to $p_{0z}$ from
Eq.~\ref{A24}, the full-pulse average of $p_{2{\mathrm{f}z}}$ is, 
\begin{equation}
\begin{tabular}{l}
$\left\langle p_{2{\mathrm{f}z}}\right\rangle =\left\langle p_{2{\mathrm{f}z}%
}\right\rangle _{\mathrm{a}}+\left\langle p_{2{\mathrm{f}z}}\right\rangle _{%
\mathrm{b}}$ \\ 
$\left\langle p_{2{\mathrm{f}z}}\right\rangle _{\mathrm{a}}=\frac{%
\left\langle p_{2{\mathrm{f}x}}^{2}\right\rangle }{2mc}$ \\ 
$\left\langle p_{2{\mathrm{f}z}}\right\rangle _{\mathrm{b}}=\frac{%
\left\langle p_{1x}^{2}\sin ^{2}\phi _{2}\right\rangle }{2mc}$%
\end{tabular}%
\ \   \label{11}
\end{equation}

We express $\left\langle p_{2{\mathrm{f}z}}\right\rangle _{\mathrm{a}}$ as $%
\left\langle p_{2{\mathrm{f}\phi }}^{2}\right\rangle \left( \cos ^{2}\phi
\right) /\left( 2mc\right) $ for a given $\phi $ particle-averaged over
solid angles $d\Omega =\left( 2\pi \sin \phi \right) d\phi $, where $p_{2{%
\mathrm{f}\phi }}$ is the post-optical momentum of electrons emitted in the $%
\phi $ direction. We determine this from the estimate to follow of the
thermalized temperature $k_{\mathrm{B}}T_{\phi }$ $=\left( 2/3\right)
\left\langle p_{2{\mathrm{f}\phi }}^{2}\right\rangle /\left( 2m\right) $ of
those electrons. We have, then,%
\begin{equation}
\left\langle p_{2{\mathrm{f}z}}\right\rangle _{\mathrm{a}}=\frac{3k_{\mathrm{%
B}}}{2c}\frac{\int\limits_{0}^{\pi /2}T_{\phi }\left( \cos ^{2}\phi \right)
\Phi _{\phi }\left( 2\pi \sin \phi \right) d\phi }{\int\limits_{0}^{\pi
/2}\Phi _{\phi }\left( 2\pi \sin \phi \right) d\phi }  \label{12}
\end{equation}%
Only $0\leq \phi \leq \pi /2$) is considered for our $\left\langle p_{2{%
\mathrm{f}z}}\right\rangle $ estimate, as discussed. We see from Eq.~\ref{09}%
.1 that this includes all the forward scattered events ($0\leq \phi _{2}\leq
\pi /2$) of interest (plus some backscatter). This is necessary since it is
not possible to distinguish between the contributions of forward scattered
electrons and some back scattered electrons from alternate optical
half-cycles. Note that, since $T_{\phi }$ is empirical, $\left\langle p_{2{%
\mathrm{f}z}}\right\rangle _{\mathrm{a}}$ includes the contribution from
unscattered electrons due to recombination or them not returing to their
parent.

The gradual variation in spectral amplitude and the\ $U$ vs.\ $\phi $
distribution of the experimental\ \cite{Okunishi08} and the theoretical \cite%
{Busuladzic08b} color maps are roughly consistent with the following
parametric representation, 
\begin{equation}
\begin{tabular}{l}
$\Phi _{\phi }=\Phi _{x}\cos ^{2}\phi +\Phi _{y}\sin ^{2}\phi $ \ \ \ $\phi
\leq \pi /2$ \\ 
$T_{\phi }=T_{x}\cos ^{2}\phi +T_{y}\sin ^{2}\phi $ \ \ \ \ \ $\phi \leq \pi
/2$%
\end{tabular}%
\ \ \ \ \ \ \   \label{13}
\end{equation}%
These forms have the necessary symmetry about the $x$-axis and $x$-$y$
plane, represent the effects of both recombination and rescatter, and have
coefficients that can be fit to our empirical $T_{x}$, $T_{+}$, and $T_{\ast
}$ results. The cited color maps show very little emission at $\phi =\pi /2$%
, so we set $\Phi _{y}=0$. This frees us to include a higher order $\Phi
_{2n}\cos ^{2n}\phi $ ($n\geq 2$) contribution to $\Phi _{\phi }$ to account
for a post-optical emission pattern more concentrated for small $\phi $ (as
suggested by the plots in more detail). This is shown below to not
significantly improve the fit, though. Neglecting this extra term, then,
implies we only need $T_{x}$ and $T_{\ast }$, since $T_{y}$ may then be
expressed in terms of them by using the fact that $T_{\ast }$ is the $\Phi
_{\phi }$-weighted average of $T_{\phi }$ over $2\pi $ sr. That is, from
Eqs.~\ref{13}, with $\Phi _{y}=0$, 
\begin{equation}
T_{\ast }=\frac{\int\limits_{0}^{\pi /2}T_{\phi }\Phi _{\phi }\left( 2\pi
\sin \phi \right) d\phi }{\int\limits_{0}^{\pi /2}\Phi _{\phi }\left( 2\pi
\sin \phi \right) d\phi }=\frac{3}{5}T_{x}+\frac{2}{5}T_{y}  \label{14}
\end{equation}%
Solving this for $T_{y}$, and substituting into Eq.~\ref{13}.2,%
\begin{equation}
\ 
\begin{tabular}{l}
$\Phi _{\phi }=\Phi _{x}\cos ^{2}\phi $ \\ 
$T_{\phi }=T_{x}\left( 1-\frac{5}{2}\sin ^{2}\phi \right) +\frac{5}{2}%
T_{\ast }\sin ^{2}\phi $%
\end{tabular}%
\ \   \label{15}
\end{equation}%
Substituting these into Eq.~\ref{12}, and integrating,%
\begin{equation}
\left\langle p_{2{\mathrm{f}z}}\right\rangle _{\mathrm{a}}=\allowbreak \frac{%
k_{\mathrm{B}}\left( 18T_{x}+45T_{\ast }\right) }{70c}  \label{16}
\end{equation}

There is a good correlation between $T_{x}$ and $T_{1}$ seen in Fig. 1 for $%
\gamma _{0}\leq 0.74$. This is not unexpected since, per assumption,
electrons recorded in the $x$-direction are not subject to the energy
enhancement due to rescatter which results in $T_{2}$. However, the near
leveling off of $T_{x}$ vs.\ $\gamma _{0}$ seen in Fig.~2 over the interval $%
0.74\leq $ $\gamma _{0}\leq 1.40$ results in a steady drop in $T_{\ast
}/T_{x}$ to near unity. This prompts the following estimates for use in Eq.~%
\ref{16}, in units of eV,%
\begin{equation}
\begin{tabular}{l}
$T_{x}=\left\{ 
\begin{tabular}{ll}
$T_{1}$ & $\gamma _{0}\leq 0.74$ \\ 
$3.10-0.362\gamma _{0}$ $\allowbreak $ & $0.74<\gamma _{0}<1.40$ \\ 
$T_{2}$ & $\gamma _{0}\geq 1.40$%
\end{tabular}%
\ \right. $ \\ 
$T_{\ast }=T_{2}$%
\end{tabular}%
\ \   \label{17}
\end{equation}%
The coefficients of the linear transition region are chosen to assure
continuity given that we have $T_{1}=2.83$ eV at $\gamma _{0}=0.74$, and $%
T_{2}=2.59$ eV at $\gamma _{0}=1.40$.

Although we have not used $T_{+}$ to refine our $\Phi _{\phi }$ profile, we
can confirm $T_{+}$ is reasonably consistent with the simpler $\Phi _{\phi }$
form of Eq.~\ref{15}.1. As the $\Phi _{\phi }$-weighted average of $T_{\phi
} $ over all angles $\phi $ in the $x$-$y$ plane instead of all solid
angles, $T_{+}$ is found by removing the $\left( 2\pi \sin \phi \right) $
terms in Eq.~\ref{14}. Plugging Eqs.~\ref{15} into the modified integral,
then, we find that\ $T_{+}$ is the weighted average $5/8$'th the way between 
$T_{x}$ and $T_{\ast }$,%
\begin{equation}
T_{+}=\allowbreak \frac{3}{8}T_{x}+\frac{5}{8}T_{\ast }  \label{18}
\end{equation}%
This compares favorably to Fig.~2, at least for the range of $\gamma _{0}$
for which there is good $T_{+}$ data.

To estimate $\left\langle p_{2{\mathrm{f}z}}\right\rangle _{\mathrm{b}}$,
meanwhile, we first estimate the particle average $\left\langle p_{2{\mathrm{%
f}z}}\right\rangle _{{\mathrm{b}\gamma }}$ of the enclosed for a given $%
\gamma $. This means averaging over $\tau _{0}$ (weighted by $W$) from $0$
to $\pi $, and over all solid angles (weighted by $\Phi _{\phi }$ of Eq.~\ref%
{15}.1) to consider. Plugging Eq.~\ref{08} into Eq.~\ref{11}.3, the result
is,%
\begin{equation}
\begin{tabular}{l}
$\left\langle p_{2{\mathrm{f}z}}\right\rangle _{{\mathrm{b}\gamma }}=$ \\ 
$\frac{\int\limits_{0}^{\pi /2}\int\limits_{\tau _{0,\min }}^{\tau _{0,\max
}}W\left( p_{1x}^{2}\sin ^{2}\phi _{2}\right) \left( \Phi _{x}\cos ^{2}\phi
\right) \left( 2\pi \sin \phi \right) d\tau _{0}d\phi }{2mc\int\limits_{0}^{%
\pi }Wd\tau _{0}\int\limits_{0}^{\pi /2}\left( \Phi _{x}\cos ^{2}\phi
\right) \left( 2\pi \sin \phi \right) d\phi }$ \\ 
$\sin \phi _{2}=\frac{\sin \tau _{1}\allowbreak \cos \phi +\sqrt{\eta
^{2}-\sin ^{2}\tau _{1}}\sin \phi }{\eta }$ $\eta =\frac{\gamma p_{1x}}{%
\sqrt{2mU_{0}}}\ $ \\ 
for $\eta ^{2}\geq \sin ^{2}\tau _{1}$ otherwise $\sin \phi _{2}=0$%
\end{tabular}%
\ \ \   \label{19}
\end{equation}

The primary $\sin \phi _{2}$ solution results from expressing Eq.~\ref{09}.3
as a quadratic in $\sin \phi _{2}$, and solving for it. The positive root
solution to the quadratic equation is chosen since it leads to the correct
limit of $\sin \phi _{2}\rightarrow \sin \phi $ as for large $%
p_{1x}\rightarrow \infty $. The condition for setting $\sin \phi _{2}=0$
corresponds to values of $\tau _{0}$ and $\eta $ for which there is no
rescatter (redundant to the $\tau _{0,\min }$ and $\tau _{0,\max }$
integration limits). Lacking data on the effect of electron-parent
recombination on\ the effective (post-optical)$\ W$, we use SFA0's Eq.~\ref%
{04}.2 (with Eq.~\ref{A13}.4, and Eq.~\ref{A27}.1) for $W$. This implies the 
$W$ integral in the denominator is twice its value from $\tau _{0}=0$ to $%
\pi /2.$

The integrals over $\phi $ have an analytic solution, resulting in,

\begin{equation}
\begin{tabular}{l}
$\left\langle p_{2{\mathrm{f}z}}\right\rangle _{{\mathrm{b}\gamma }}=\left(
20mc\int\limits_{0}^{\pi /2}Wd\tau _{0}\right) ^{-1}\int\limits_{\tau
_{0,\min }}^{\tau _{0,\max }}\eta ^{-2}p_{1x}^{2}W$ \\ 
$\times \left( \sin ^{2}\tau _{1}\allowbreak +4\sqrt{\eta ^{2}-\sin ^{2}\tau
_{1}}\sin \tau _{1}\allowbreak \allowbreak +2\eta ^{2}\allowbreak \right)
d\tau _{0}$%
\end{tabular}%
\ \   \label{20}
\end{equation}%
with the $\sin \phi _{2}=0$ condition of Eq.~\ref{19}.4 still applicable.
The remaining integral is solved numerically, where Eq.~\ref{08} and the
Newton's method solution plotted in Fig.~3a are used for $p_{1x}$ and $\tau
_{1}$, respectively. $U_{2{\mathrm{f}z}}=\left\langle p_{2{\mathrm{f}z}%
}\right\rangle ^{2}/(2m)$, based on the use of this approximation in Eqs.~%
\ref{11}, and, for comparison, SFA0's $U_{0{\mathrm{f}z}}=\left\langle p_{{%
\mathrm{f}z}}\right\rangle ^{2}/(2m)$, based on Eq.~\ref{05} with $%
T_{x}=T_{0}$ from Eq.~\ref{03}.3, are plotted in Fig.~1. This representation
is chosen to conform to the plot's ordinate eV units.

Approximating $\left\langle p_{2{\mathrm{f}z}}\right\rangle _{\mathrm{b}}$
by $\left\langle p_{2{\mathrm{f}z}}\right\rangle _{\mathrm{b}\gamma _{0}}$
is reasonably accurate if there is a high enough $\left\langle
W\right\rangle _{\gamma }$ dependence on $\gamma $. To establish when such
is the case, we estimate the error for the case of an optical pulse with
both a Gaussian radial density and temporal profile, and with $\left\langle
W\right\rangle _{\gamma }$ obeying a power law in $I$. The sample length
over which electrons are collected in the referenced data is not specified.
If we assume that it is small compared to the Rayleigh range \cite{Siegman86}%
, $\left\langle p_{2{\mathrm{f}z}}\right\rangle _{\mathrm{b}}$ is the
average of $\left\langle p_{2{\mathrm{f}z}}\right\rangle _{{\mathrm{b}\gamma 
}}$ over the cross section and time of the laser pulse. Assuming weak
ionization, 
\begin{equation}
\left\langle p_{2{\mathrm{f}z}}\right\rangle _{\mathrm{b}}=\frac{%
\int\limits_{0}^{+\infty }r\int\limits_{-\infty }^{+\infty }\left\langle p_{2%
{\mathrm{f}z}}\right\rangle _{{\mathrm{b}\gamma }}\left\langle
W\right\rangle _{\gamma }dtdr}{\int\limits_{0}^{+\infty
}r\int\limits_{-\infty }^{+\infty }\left\langle W\right\rangle _{\gamma }dtdr%
}  \label{24}
\end{equation}%
$\left\langle p_{2{\mathrm{f}z}}\right\rangle _{\mathrm{b}\gamma
_{0}}>\left\langle p_{2{\mathrm{f}z}}\right\rangle _{\mathrm{b}}$ since $%
\left\langle p_{2{\mathrm{f}z}}\right\rangle _{\mathrm{b}\gamma }$ is
characteristic of a given $\gamma $ (set equal to $\gamma _{0}$ in the
inequality), while $\left\langle p_{2{\mathrm{f}z}}\right\rangle _{\mathrm{b}%
}$ refers to the average of a pulse that only \emph{peaks} in intensity at $%
\gamma _{0}$. We define, then,

\begin{equation}
\alpha \left( \gamma _{0}\right) =\frac{\left\langle p_{2{\mathrm{f}z}%
}\right\rangle _{\mathrm{b}}}{\left\langle p_{2{\mathrm{f}z}}\right\rangle _{%
\mathrm{b}\gamma _{0}}}{<1}  \label{25}
\end{equation}

If $\left\langle W\right\rangle _{\gamma }$ obeys a power law in $I$ with
constant real coefficient $\mu $, and $I$ is a fixed normalized
spatiotemporal distribution multiplied by $I_{0}$ (no refraction), then the
number of electrons released in a fixed sample volume is proportional to $%
I_{0}^{\mu }$ \emph{provided} ionization results in negligible depletion of
neutral O$_{2}$ molecules. $\mu =8$ for O$_{2}$ in the multiphoton regime ($%
\gamma \gg 1$) \cite{Sharma18}. However, the goodness of fit to a straight
line of Guo et al.'s, \cite{Guo98} unscaled log-log plot of O$_{2}^{+}$
count vs.\ $I_{0}$ over a significant range of $I_{0}$ shows that $%
\left\langle W\right\rangle _{\gamma }\propto I^{\mu \left( \gamma
_{0}\right) }$ is sufficiently accurate over a range of $I\lesssim I_{0}$
for us to treat it locally as a power law, at least down to $\gamma _{0}\sim
1$. That is, $\mu \left( \gamma _{0}\right) $ is a decreasing function of $%
\gamma _{0}$, but treated as constant relative to $I$ for a given $\gamma
_{0}$. From Fig.~2, we see that $\left\langle p_{2{\mathrm{f}z}%
}\right\rangle _{{\mathrm{b}\gamma }}$ varies much more linearly with $I$
than $\left\langle W\right\rangle _{\gamma }$. Given that only $\left\langle
p_{2{\mathrm{f}z}}\right\rangle _{{\mathrm{b}\gamma }}$ values for $\gamma $
near $\gamma _{0}$ contribute significantly to $\left\langle p_{2{\mathrm{f}z%
}}\right\rangle _{\mathrm{b}}$, then, a first order (linear) extrapolation
of $\left\langle p_{2{\mathrm{f}z}}\right\rangle _{{\mathrm{b}\gamma }}$
vs.\ $I$ \ from its value at $I=I_{0}$ (i.e. $\gamma =\gamma _{0}$) in the
Eq.~\ref{24} is sufficient.

The above two approximations imply the power law and Eq.~\ref{25} may be
respectively expressed as, 
\begin{equation}
\begin{tabular}{l}
$\left\langle W\right\rangle _{\gamma }=\sigma _{\mu }\left( \gamma
_{0}\right) I^{\mu \left( \gamma _{0}\right) }$ \ $\ \ \ \ \ \ I=\frac{%
\omega ^{2}mU_{0}c\epsilon _{0}}{e^{2}\gamma ^{2}}$ \\ 
$\left\langle p_{2{\mathrm{f}z}}\right\rangle _{{\mathrm{b}\gamma }%
}=\left\langle p_{2{\mathrm{f}z}}\right\rangle _{{\mathrm{b}\gamma _{0}}}+%
\frac{\partial \left\langle p_{2{\mathrm{f}z}}\right\rangle _{{\mathrm{b}%
\gamma _{0}}}}{\partial I_{0}}\left( I-I_{0}\right) $%
\end{tabular}%
\ \   \label{26}
\end{equation}%
where $\sigma _{\mu }\left( \gamma _{0}\right) $ is a $\gamma _{0}$
dependent proportionality constant (which will not be needed). Substituting
the above expressions into Eq.~\ref{24}, the resulting expression for $%
\left\langle p_{2{\mathrm{f}z}}\right\rangle _{\mathrm{b}}$ into Eq.~\ref{25}%
, and solving that for $\gamma =\gamma _{0}$, 
\begin{equation}
\begin{tabular}{l}
$\alpha \left( \gamma _{0}\right) =\frac{\int\limits_{0}^{+\infty
}r\int\limits_{-\infty }^{+\infty }\left( \left\langle p_{2{\mathrm{f}z}%
}\right\rangle _{{\mathrm{b}\gamma _{0}}}+\frac{\partial \left\langle p_{2{%
\mathrm{f}z}}\right\rangle _{{\mathrm{b}\gamma _{0}}}}{\partial I_{0}}\left(
I-I_{0}\right) \right) I^{\mu \left( \gamma _{0}\right) }dtdr}{\left\langle
p_{2{\mathrm{f}z}}\right\rangle _{\mathrm{b}\gamma
_{0}}\int\limits_{0}^{+\infty }r\int\limits_{-\infty }^{+\infty }I^{\mu
\left( \gamma _{0}\right) }dtdr}$%
\end{tabular}
\label{27}
\end{equation}%
If the $I$ profile is Gaussian in both $r$ and $t$,

\begin{equation}
\begin{tabular}{l}
$I=I_{0}\exp \left( -\frac{t^{2}}{t_{\mathrm{L}}^{2}}-\frac{2r^{2}}{w_{0}^{2}%
}\right) $ \\ 
$\alpha \left( \gamma _{0}\right) =1-\left[ \frac{I_{0}}{\left\langle p_{2{%
\mathrm{f}z}}\right\rangle _{{\mathrm{b}\gamma _{0}}}}\frac{\partial
\left\langle p_{2{\mathrm{f}z}}\right\rangle _{{\mathrm{b}\gamma _{0}}}}{%
\partial I_{0}}\right] $ \\ 
$\times \left( 1-\left( \frac{\mu \left( \gamma _{0}\right) }{1+\mu \left(
\gamma _{0}\right) }\right) ^{3/2}\right) $%
\end{tabular}%
\ \   \label{28}
\end{equation}%
where $t_{\mathrm{L}}$ and $w_{0}$ are the $e^{-1}$ temporal half-width and $%
e^{-2}$ radial half-width of $I$, respectively.

For $\gamma _{0}=1$ ($I_{0}=101$ TW/cm$^{2}$), the plot of Guo, et al, \cite%
{Guo98} shows $\mu \left( \gamma _{0}\right) =4.43$. Meanwhile, $%
\left\langle p_{2{\mathrm{f}z}}\right\rangle _{\mathrm{a}}=1.66\times
10^{-27}$ kg-m/s from Eq.~\ref{16} with $T_{x}=\allowbreak 2.74$ eV, and $%
T_{\ast }=3.9$ eV, and $\left\langle p_{2{\mathrm{f}z}}\right\rangle _{%
\mathrm{b}\gamma _{0}}=0.80\times 10^{-27}$ kg-m/s from the numerical
solution to Eq.~\ref{20} for $\gamma _{0}=1$. The term in square brackets of
the Eq.~\ref{28}.2 is $0.90$, from finite differences of calculated values.
From these, $\alpha \left( \gamma _{0}\right) =\allowbreak 0.76\allowbreak $%
, so $\left\langle p_{2{\mathrm{f}z}}\right\rangle =\left\langle p_{2{%
\mathrm{f}z}}\right\rangle _{\mathrm{a}}+\alpha \left( \gamma _{0}\right)
\left\langle p_{2{\mathrm{f}z}}\right\rangle _{\mathrm{b}\gamma
_{0}}=2.27\times 10^{-27}$ kg-m/s. This is an 8\% reduction relative to $%
\left\langle p_{2{\mathrm{f}z}}\right\rangle _{\mathrm{a}}+\left\langle p_{2{%
\mathrm{f}z}}\right\rangle _{\mathrm{b}\gamma _{0}}$. One sees from Eq.~\ref%
{28}.2 that the correction factor drops even further for $\gamma _{0}>1$ due
to higher $\mu \left( \gamma _{0}\right) $. The correction is more
significant for $\gamma _{0}<1$.

The above $\alpha \left( \gamma _{0}\right) $ correction term for $%
\left\langle p_{2{\mathrm{f}z}}\right\rangle _{\mathrm{b}\gamma _{0}}$,
recall, assumes that ionization is weak. The discrepancy between $%
\left\langle p_{2{\mathrm{f}z}}\right\rangle _{\mathrm{b}\gamma _{0}}$ and $%
\left\langle p_{2{\mathrm{f}z}}\right\rangle _{\mathrm{b}\gamma }$ is
greater if this is not the case. A recent estimate of the absolute $\mathrm{%
\ }\left\langle W\right\rangle _{\gamma }$ vs.~$I_{0}$ of O$_{2}$ \cite%
{Ruden25Qmodn} used to determine the post-optical percent of O$_{2}$
ionization vs.~$I_{0}$ for a 800 nm USPL laser with a full-width at half-max
of 100 fs \cite{Ruden25QmodS} found that there is 50\% depletion at $%
I_{0}=1.34\times 10^{14}$ W/cm$^{2}$ ($\gamma _{0}=0.87$). This implies
significant error in our model, as it applies to the $T_{x}$ for $\gamma
_{0}<1$ (fit to data with this FWHM, per Table 1). \ Taking $T_{\ast }=T_{2}$%
, though, based on data from a much shorter pulse, is not as significantly
compromised by this error, as long as $\gamma _{0}\approx 1$.

\section{Summary and conclusions}

A semi-empirical model is presented for the thermalized temperature $T$ and
mean momentum in the direction of laser propagation $\left\langle p_{{%
\mathrm{f}z}}\right\rangle $ of electrons released from O$_{2}$ molecules
after exposure to a focused $800$ nm ultrashort pulsed laser (USPL) pulse as
a function of\ peak laser intensity $I_{0}$. The purpose is to provide
initial conditions for simulations that assume electron thermalization and
an initial current density.

To achieve this, a published model based on the most probable tunnel path of
a strong field approximation (SFA0 in the main text) is reformulated. This
model yields the instantaneous ionization rate $W=C_{\gamma }C_{0}\exp (-2G_{%
\mathrm{c}}/\hbar )$ as a function of ionization time $t_{0}$ of O$_{2}$
exposed to an $800$ nm USPL pulse. Appendix A develops SFA0, which addresses
both linear ($\varepsilon =0$) and circular ($\varepsilon =1$)\
polarization. However, the focus narrows to $\varepsilon =0$ when the model
is fit to empirical spectra, due to limited data for $\varepsilon =1$.$\ G_{%
\mathrm{c}}$ is the global minimum of $G$ w.r.t.\ tunnel time $t_{\mathrm{i}%
} $ (at $t_{\mathrm{i}}=t_{\mathrm{ic}}$), and plotted in Fig.~A2. $G_{%
\mathrm{c}}$ is the local minimum of $G$ (Eq.~\ref{A13}.1) w.r.t.\ $t_{%
\mathrm{i}}$ for $\varepsilon =1$, and its value at the smallest possible $%
t_{\mathrm{i}}$ for $\varepsilon =0$ (Eq.~\ref{A14}.1). $C_{0}$ vs.\ $t_{0}$
for a range of Keldysh parameters $\gamma $ (Eq.~\ref{A13}.8) is determined
by Eqs.~\ref{A27} and plotted in Fig.~A4. $C_{\gamma }$ vs.\ $\gamma $ is an
overall scaling factor independent of $t_{0}$ determined empirically based
on $W$'s optical cycle average $\left\langle W\right\rangle _{\gamma }$ vs.\ 
$\gamma $. It is obtained by inverting published full-pulse ionization
product data vs.\ peak laser intensity $I_{0}$ in a complementary paper \cite%
{Ruden25Qmodn}, but not needed for this one.

SFA0 also provides post-optical (immediately after the optical pulse) and
residual (upon ionization) electron momenta $\mathbf{p}_{{\mathrm{f}}}$ and $%
\mathbf{p}_{0}$, respectively. Their components orthogonal to the direction
of laser propagation ($z$) are determined by Eqs.~\ref{A15} and Eqs.~\ref%
{A18}, respectively. The model is extended relative to its published form to
include the $z$ component by Eqs.~\ref{A24}. These results are plotted in
Fig.~A3.

The electron temperature $T_{0}$, resulting from SFA0 after thermalization
of the kinetic energy spectrum $S_{0}\left( U\right) $ is determined using
Eqs.~\ref{03}. For $\varepsilon =0$, $S_{0}\left( U\right) $ serves as a
fitting function for published empirical spectra with two phenomenological
parameters: (1) a ceiling to suppress an unobserved surge in $S_{0}\left(
U\right) $ for low energies (yielding $S_{1}\left( U\right) $ for model
SFA1), and (2) an energy multiplier to account for the observed increase in
temperature (resulting in $S_{2}\left( U\right) $ for model SFA2). These
parameters are necessary because electrons can recombine with or rescatter
off their parent ion within an optical cycle. This process selectively
removes low-energy electrons and accelerates electrons released near the
peak of the optical electric field to much higher post-optical energies than
would otherwise occur. Examples of theoretical and empirical spectra are
overlaid in Fig.~1. Empirically inferred thermalized temperatures $T$ from
published spectra recorded in the ($\mathbf{E}$ field) $x$-direction for $%
\varepsilon =0$ (with subscript $x$), averaged over $2\pi $ rad in the $x$-$%
y $ plane (with subscript $+$), and averaged over $4\pi $ steridians (with
subscript $\ast $) are overlaid with $T_{0}$ (from $S_{0}\left( U\right) $), 
$T_{1}$ (from $S_{1}\left( U\right) $), and $T_{2}$ (from $S_{2}\left(
U\right) $) in Fig.~2.

Section III calculates the effect of electron-parent ion recombination and
rescatter on the electrons' mean post-optical axial momentum $\left\langle
p_{2{\mathrm{f}z}}\right\rangle $ (Eq.~\ref{11}.1), based on SFA2 and
classical kinematics applied to an idealized scatter pattern (Eqs.~\ref{13}%
). The result is plotted in Fig.~2, and compared to SFA0. Although the SFA2
total temperature $T_{\ast }$ estimate is most accurate for $0.82\leq $ $%
\gamma _{0}\leq 1.30$ (since it is based on an interpolation between $T$
data points in this range), $\left\langle p_{2{\mathrm{f}z}}\right\rangle $
is expected to have significant error due to its use of $\left\langle p_{2{%
\mathrm{f}z}}\right\rangle _{\mathrm{b}\gamma _{0}}$ to approximate its $%
\left\langle p_{2{\mathrm{f}z}}\right\rangle _{\mathrm{b}}$ contribution,
especially for $\gamma <1$, where neutral O$_{2}$ depletion is significant
for the long pulse (100 fs) $T_{x}$ data fit to.

On a final note, since the empirical spectra used for the fits result from a
full optical pulse of \emph{peak} intensity $I_{0}$, SFA2 is interpreted as
representing the spatiotemporal average effects of one. However, we can
adapt the model for use with spatiotemporally \emph{resolved} models (where
local properties are needed). To do so, the procedure\ in Sec.~III for
estimating the factor $\alpha \left( \gamma _{0}\right) $ by which $%
\left\langle p_{2{\mathrm{f}z}}\right\rangle _{\mathrm{b}}$ (full-pulse
average) is smaller than $\left\langle p_{2{\mathrm{f}z}}\right\rangle _{%
\mathrm{b}\gamma _{0}}$ (local average, but with $\gamma =\gamma _{0}$) may
be reversed by setting $Q_{\gamma }\left( I\right) =Q\left( I\right) /\alpha
_{Q}\left( \gamma \right) $, where $Q_{\gamma }\left( I\right) $ is a local
property and $Q\left( I_{0}\right) $ is its full-pulse average (with $%
I_{0}=I $ in the equality). Conversion factor $\alpha _{Q}\left( \gamma
\right) $ is assumed to vary slowly enough with $I$ that it may be treated
as a constant of integration equal to $\alpha _{Q}\left( \gamma _{0}\right) $
for the purpose of averaging $Q_{\gamma }\left( I\right) $ over a full pulse
of peak intensity $I_{0}$. Approximating $Q\left( I_{0}^{\prime }\right) $
in the vicinity of $I_{0}^{\prime }=I_{0}$ by,%
\begin{equation}
Q\left( I_{0}^{\prime }\right) =Q\left( I_{0}\right) +\frac{Q\left(
I_{0}\right) }{\partial I_{0}}\left( I_{0}^{\prime }-I_{0}\right)  \label{29}
\end{equation}%
One then finds,%
\begin{equation}
\begin{tabular}{l}
$\alpha _{Q}\left( \gamma \right) =1-\frac{I}{Q\left( I\right) }\frac{%
\partial Q\left( I\right) }{\partial I}\left( 1-\left( \frac{\mu \left(
\gamma \right) }{1+\mu \left( \gamma \right) }\right) ^{3/2}\right) $%
\end{tabular}
\label{30}
\end{equation}%
Though similar in form to Eq.~\ref{28}.2, the roles of local vs.~full-pulse
properties are reversed. If greater accuracy is required, inverting $Q\left(
I_{0}\right) $ to determine $Q_{\gamma }\left( I\right) $ may be
accomplished by the generalized Newton method presented in the complementary
paper \cite{Ruden25Qmodn}.

\appendix

\section{The Strong Field Approximation's most probable ionization path}

\subsection{SFA for $G$}

This appendix presents a reformulation of and expansion on the strong field
approximation (SFA) of Li, et al.\ \cite{Li17} (referred to here as
\textquotedblleft Li\textquotedblright ) and Luo, et al.\ \cite{Luo19}
(\textquotedblleft Luo\textquotedblright ), based on the most probable
tunnel path for molecular ionization and subsequent electron acceleration.
The vector potential $\mathbf{A}$ and electric field $\mathbf{E}$ to which
an O$_{2}$ molecule is exposed are assumed to be,%
\begin{equation}
\begin{tabular}{l}
$\mathbf{A}=-\frac{\mathcal{E}\left( t-z/c\right) }{\omega }\left( \sin
\left( \omega t-kz\right) \mathbf{\hat{e}}_{x}\right. $ \\ 
$\left. +\varepsilon \cos \left( \omega t-kz\right) \mathbf{\hat{e}}%
_{y}\right) $ \\ 
$\mathbf{E}=\mathcal{E}\left( t-z/c\right) \left( \cos \left( \omega
t-kz\right) \mathbf{\hat{e}}_{x}\right. $ \\ 
$\left. +\varepsilon \sin \left( \omega t-kz\right) \mathbf{\hat{e}}%
_{y}\right) $%
\end{tabular}%
\ \   \label{A01}
\end{equation}%
where $c$ is the speed of light, and $k=\omega /c$. This describes an
elliptically polarized optical pulse with central angular frequency $\omega $%
, ellipticity $\varepsilon $, and an $\mathbf{E}$ envelope $\mathcal{E}%
\left( t\right) $ propagating in the $\mathbf{\hat{e}}_{z}$ ($z$ unit
vector) direction at a speed $c$. $\varepsilon =0$ for linear polarization
and $\varepsilon =1$ for circular polarization. The laser intensity is\emph{%
\ }$I\left( t\right) =$ $c\epsilon _{0}\mathcal{E}^{2}\left( t\right) /2$,
where $\epsilon _{0}$ is the permittivity of free space. \ The wavelength is 
$2\pi c/\omega =800$ nm for the laser to which the model is applied. The
full width at half maximum of $I\left( t\right) $ is assumed to be much
greater than $2\pi /\omega $, so $\mathcal{E}$ is reasonably treated as
constant for intracycle operations.

The nonrelativistic classical Lagrangian of an electron in an
electromagnetic (EM)\ field is \cite{Landau75},%
\begin{equation}
\mathcal{L}=\frac{m}{2}\mathbf{\dot{x}}^{2}-e\mathbf{\dot{x}}\cdot \mathbf{A}%
+e\phi \left( \mathbf{x},t\right) {\mathrm{\ \ \ }\mathbf{\dot{x}=}\frac{d%
\mathbf{x}}{dt}}  \label{A02}
\end{equation}%
where $\phi $ is the electric scalar potential. This is the basis of the
Hamiltonian used for the Green function solution to the Schr\"{o}dinger
equation \cite{Feynman71} that the SFA is an approximation of \cite{Popov05}%
. Operation of the Euler-Lorentz equation \cite{LL76} on this expression
gives us the Lorentz plus Coulomb force due to the remaining atomic
structure acting on the electron,%
\begin{equation}
\begin{tabular}{l}
$\frac{d}{dt}\left( \frac{\partial \mathcal{L}}{\partial \mathbf{\dot{x}}}%
\right) -\frac{\partial \mathcal{L}}{\partial \mathbf{x}}=0$ \ \ \ \ \ \ \ \
\ \ $\mathbf{p=}m\mathbf{\dot{x}}$ \\ 
$\frac{d}{dt}\mathbf{p}=e\frac{\partial \mathbf{A}}{\partial t}-\left[ \frac{%
e}{m}\mathbf{p}\times \left( \nabla \times \mathbf{A}\right) \mathbf{+}e%
\frac{\partial \phi }{\partial \mathbf{x}}\right] $%
\end{tabular}
\label{A03}
\end{equation}%
These expressions are used by the SFA to classically describe the motion of
the electron after being released into the continuum at $z=0$ and $t=t_{0}$.
They are also analytically continued to quasiclassically describe electron
tunneling, based on the imaginary time method \cite{Popov05}. This results
in complex expressions for action and momentum, based on integration over
complex\emph{\ }time $t$ from $t_{\mathrm{s}}=t_{0}+it_{\mathrm{i}}$ to $%
t_{0}$ along a path\ parallel to the imaginary time axis, representing
tunneling between the ground and continuum states \cite{Li17}. $t_{\mathrm{i}%
}$ is a free parameter representing the stochastic nature of location and
momentum of the ground state. It is referred to as the \textquotedblleft
tunnel time\textquotedblright\ since the $i\omega t_{\mathrm{i}}$
contribution to the sine and cosine arguments of Eqs.~\ref{A01} along the
tunnel path implies that $t_{\mathrm{i}}$ is the $e$-fold time that the
external EM field couples to the electron over this path.

We neglect the term in square brackets in Eq.~\ref{A03}.2 (the equation on
line 2 of Eqs.~\ref{A03}) that results from the electron's momentum vector
crossing the magnetic field and the Coulomb interaction with the remaining
atomic structure for an initial first order treatment. The Coulomb
interaction is assumed to be localized near the origin and characterized by
ionization potential $U_{0}$ for this. Integrating Eq.~\ref{A03}.2 from $%
t_{0}$\ to variable $t$ (complex during tunneling) we have, then, to first
order, the momentum vector in the cylindrical radius $r$ direction
transverse to the $z$-axis, 
\begin{equation}
\mathbf{p}_{r}=\mathbf{p}_{{\mathrm{f}r}}+e\mathbf{A}\ \ \   \label{A04}
\end{equation}
$\mathbf{p}_{{\mathrm{f}r}}$ here is the final value of $\mathbf{p}_{r}$
after optical pulse passage for a given $t_{0}$. This follows from $\mathbf{%
A=0}$ for $t\rightarrow\infty$ due to the finite pulse duration.

Our first order approximation is equivalent to Eq.~\ref{A02} being
approximated by the Lagrangian of an electron in a uniform $\mathbf{E}$
field \cite{Popov05}, 
\begin{equation}
\mathcal{L}=\frac{p^{2}}{2m}+e\mathbf{E\cdot x}  \label{A05}
\end{equation}%
Given this, Popov 2005 \cite{Popov05} derives the action $S$ (Popov
Eq.~2.11) and its saddle equation constraint (Popov Eq.~2.9) that serve as
the basis for the SFA from the Green function integral representation of the
solution to the Schr\"{o}dinger Equation (Popov Eq.~2.3). Given Eq.~\ref{A04}%
, $S$ is (Luo Eq.~3 \cite{Luo19}), 
\begin{equation}
S=-\int\limits_{t_{\mathrm{s}}}^{t_{0}}\left( \frac{1}{2m}\left( \mathbf{p}_{%
{\mathrm{f}r}}+e\mathbf{A}_{0}\left( t\right) \right) ^{2}+U_{0}\right) dt
\label{A06}
\end{equation}%
where $\mathbf{A}_{0}\left( t\right) $ is defined as $\mathbf{A}$ at $z=0$
and time $t$, and the integration is along the complex tunnel path defined
above. The saddle equation (Luo Eq.~10 \cite{Luo19}), meanwhile, is,%
\begin{equation}
-\frac{dS}{dt_{\text{s}}}=\frac{1}{2m}\left( \mathbf{p}_{{\mathrm{f}r}}+e%
\mathbf{A}_{0}\left( t_{\mathrm{s}}\right) \right) ^{2}+U_{0}=0  \label{A07}
\end{equation}

Integrating Eq.~\ref{A06} and separating out the imaginary component, we
obtain (Luo Eq.~5 \cite{Luo19}), 
\begin{equation}
\begin{tabular}{r}
$G=$Im$S=\left( \frac{p_{{\mathrm{f}x}}^{2}+p_{{\mathrm{f}y}}^{2}}{2m}+U_{0}+%
\frac{e^{2}\left( 1+\varepsilon ^{2}\right) \mathcal{E}^{2}}{4m\omega ^{2}}%
\right) t_{\mathrm{i}}$ \\ 
$-\frac{e\mathcal{E}\sinh \omega t_{\mathrm{i}}}{m\omega ^{2}}\eta -\frac{%
\left( 1-\varepsilon ^{2}\right) e^{2}\mathcal{E}^{2}\cos 2\omega t_{0}\sinh
2\omega t_{\mathrm{i}}}{8m\omega ^{3}}$%
\end{tabular}
\label{A08}
\end{equation}%
where the ionization rate is (Luo Eq.~2 \cite{Luo19}),$\ $%
\begin{equation}
W=C_{\gamma }C_{0}\exp (-2G/\hbar )  \label{A09}
\end{equation}%
$\eta $ is defined by Eq.~\ref{A11}.1 below as part of a change of
variables. $C_{\gamma }$ is a laser intensity-dependent scaling factor (not
needed here), and $C_{0}$ is the subject of Sec.~A4.

At this point we \ depart from Luo's approach \cite{Luo19} for the reasons
discussed in Appx.~A.5. Substituting the components of $\mathbf{A}$ from Eq.~%
\ref{A01}.1, into Eq.~\ref{A07}, and zeroing both real and imaginary parts,
respectively, we have at $z=0$,%
\begin{equation}
\begin{tabular}{l}
$\frac{p_{{\mathrm{f}x}}^{2}+p_{{\mathrm{f}y}}^{2}}{2m}+U_{0}=\frac{e%
\mathcal{E}\eta \cosh \omega t_{\mathrm{i}}}{m\omega }$ \\ 
$+\frac{e^{2}\mathcal{E}^{2}\left( \left( 1-\varepsilon ^{2}\right) \cos
2\omega t_{0}\cosh 2\omega t_{\mathrm{i}}-\varepsilon ^{2}-1\right) }{%
4m\omega ^{2}}$ \\ 
$\xi =\frac{\allowbreak e\mathcal{E}\left( 1-\allowbreak \varepsilon
^{2}\right) \sin 2\omega t_{0}\cosh \omega t_{\mathrm{i}}}{2\omega }$%
\end{tabular}%
\ \   \label{A10}
\end{equation}%
where the following transformation and its inverse are,%
\begin{equation}
\begin{tabular}{l}
$\eta =p_{{\mathrm{f}x}}\sin \omega t_{0}-\varepsilon p_{{\mathrm{f}y}}\cos
\omega t_{0}$ \\ 
$\xi =p_{{\mathrm{f}x}}\cos \omega t_{0}+\allowbreak \varepsilon p_{{\mathrm{%
f}y}}\sin \omega t_{0}$ \\ 
If $\varepsilon \neq 0\ $then $p_{{\mathrm{f}x}}=\eta \sin \omega t_{0}+\xi
\cos \omega t_{0}$ \\ 
and $p_{{\mathrm{f}y}}=\left( \xi \sin \omega t_{0}-\eta \cos \omega
t_{0}\right) /\varepsilon $ \\ 
If $\varepsilon =0\ $then$\ p_{{\mathrm{f}x}}=\eta /\sin \omega t_{0}$ and $%
p_{{\mathrm{f}y}}=0$ \ 
\end{tabular}%
\ \   \label{A11}
\end{equation}%
The inverses here are determined by Gaussian elimination, except for $%
\varepsilon =0$, where the result is indeterminate. $p_{{\mathrm{f}x}}$ in
that case follows directly from the definition of $\eta $, and $p_{{\mathrm{f%
}y}}=0$ is assigned for $\varepsilon =0$ from symmetry. Here and henceforth,
the limiting case of $\varepsilon =0$ will be carried through separately.

An expression for $\eta$ in terms of $t_{\mathrm{i}}$ which satisfies Eq.~%
\ref{A07} is found by substituting $\xi$ in Eq.~\ref{A10}.3 into Eq.~\ref%
{A11}.3 and Eq.~\ref{A11}.4 for $\varepsilon\neq0$, and then substituting
the expressions for $p_{{\mathrm{f}x}}$ and $p_{{\mathrm{f}y}}$ for all $%
\varepsilon$ values into Eq.~\ref{A10}.1. This leads to the following
quadratics in $\eta$, with solutions for $\varepsilon=1$ and $\varepsilon=0$%
, 
\begin{equation}
\begin{tabular}{l}
$\varepsilon\neq0$:\ \ $0=4\omega^{2}\eta^{2}\left( \cos^{2}\omega
t_{0}+\varepsilon^{2}\sin^{2}\omega t_{0}\right) $ \\ 
$-2\omega e\mathcal{E}\eta\left( \left( 1-\varepsilon^{2}\right)
^{2}\sin^{2}2\omega t_{0}\allowbreak+4\varepsilon^{2}\right) \cosh\omega t_{%
\mathrm{i}}$ \\ 
$+e^{2}\mathcal{E}^{2}\left( \left( 1-\varepsilon^{2}\right)
^{2}\allowbreak\sin^{2}2\omega t_{0}+\allowbreak4\varepsilon^{2}\right) $ \\ 
$\times\left( \sin^{2}\omega t_{0}+\varepsilon^{2}\cos^{2}\omega
t_{0}\right) \cosh^{2}\omega t_{\mathrm{i}}$ \\ 
$-4\varepsilon^{2}e^{2}\mathcal{E}^{2}\left( \allowbreak\cos^{2}\omega
t_{0}+\varepsilon^{2}\sin^{2}\omega t_{0}\right) \sinh^{2}\omega t_{\mathrm{i%
}}$ \\ 
$+\allowbreak8\varepsilon^{2}\omega^{2}mU_{0}$ \\ 
$\varepsilon=1$: $\allowbreak\allowbreak$ \ $0=\eta^{2}\omega^{2}-2\omega e%
\mathcal{E}\eta\cosh\omega t_{\mathrm{i}}$ \\ 
$+e^{2}\mathcal{E}^{2}+2\allowbreak\omega^{2}mU_{0}$ \ \ \ \ \ \ \ \ \ $\eta
=\frac{e\mathcal{E}\cosh\omega t_{\mathrm{i}}-R_{1}}{\omega}$ \\ 
$\allowbreak R_{1}=\sqrt{e^{2}\mathcal{E}^{2}\sinh^{2}\omega t_{\mathrm{i}%
}-2m\omega^{2}U_{0}}$ \\ 
$\varepsilon=0$:\ $0=2\eta^{2}\omega^{2}-4\omega e\mathcal{E}\eta\sin
^{2}\omega t_{0}\cosh\omega t_{\mathrm{i}}$ \\ 
$-e^{2}\mathcal{E}^{2}\left( \cos2\omega t_{0}\cosh2\omega t_{\mathrm{i}%
}-1\right) \sin^{2}\omega t_{0}\allowbreak$ \\ 
$+4\omega^{2}mU_{0}\sin^{2}\omega t_{0}$ \\ 
$\eta=\frac{\left( e\mathcal{E}\sin^{2}\omega t_{0}\cosh\omega t_{\mathrm{i}%
}+R_{0}\left\vert \sin\omega t_{0}\right\vert \right) }{\omega}$ \\ 
$R_{0}=\sqrt{e^{2}\mathcal{E}^{2}\cos^{2}\omega t_{0}\sinh^{2}\omega t_{%
\mathrm{i}}-2m\omega^{2}U_{0}}$%
\end{tabular}
\label{A12}
\end{equation}

We now confine our attention to $\varepsilon=1$ and $\varepsilon=0$. The
physically meaningful solutions to $\eta$ of the two provided by the
quadratic equation (with $\pm$ the root) are determined by substituting the
r.h.s.\ of Eq.~\ref{A10}.1 into the term to which it is equal to in Eq.~\ref%
{A08}, substituting the values of $\eta$ from Eqs.~\ref{A12} for our two $%
\varepsilon$ cases into the result, then comparing the result to Eq.~\ref%
{A08} for select numerical values of the parameters. The absolute value
taken of the $\sin\omega t_{0}$ term multiplying $R_{0}$ above avoids the
need to specify a change in the sign of the root for $\pi<\omega t_{0}<2\pi$%
. Simplifying Eq.~\ref{A08} so obtained and switching to normalized time, we
have, 
\begin{equation}
\begin{tabular}{l}
$\varepsilon=1$:\ \ $\frac{2G}{\hbar}=\frac{U_{0}}{\gamma^{2}\hbar\omega}$
\\ 
$\times\left( \left( 4\cosh^{2}\tau_{\mathrm{i}}-4\sqrt{\sinh^{2}\tau_{%
\mathrm{i}}-\gamma^{2}}\cosh\tau_{\mathrm{i}}\right) \tau _{\mathrm{i}%
}\right. $ \\ 
$\left. -2\sinh2\tau_{i}+4\sqrt{\sinh^{2}\tau_{\mathrm{i}}-\gamma^{2}}%
\sinh\tau_{\mathrm{i}}\right) $ \\ 
$\varepsilon=0$:\ \ $\frac{2G}{\hbar}=\frac{U_{0}}{\hbar\omega}$ \\ 
$\times\left( \frac{\left( \allowbreak\cosh2\tau_{\mathrm{i}}-\cos
2\tau_{0}+1+4R_{3}\left\vert \sin\tau_{0}\right\vert \cosh\tau_{\mathrm{i}%
}\right) \tau_{\mathrm{i}}}{\gamma^{2}}\right. $ \\ 
$\left. -\frac{\left( 1+2\sin^{2}\tau_{0}\right) \sinh2\tau_{\mathrm{i}%
}+8R_{3}\left\vert \sin\tau_{0}\right\vert \sinh\tau_{\mathrm{i}}}{%
2\gamma^{2}}\right) $ \\ 
$R_{3}=\sqrt{\cos^{2}\tau_{0}\sinh^{2}\tau_{\mathrm{i}}-\gamma^{2}}$ \\ 
$\gamma=\frac{\omega\sqrt{2mU_{0}}}{e\mathcal{E}}{\ \ \ \ \ \ }%
\tau_{0}=\omega t_{0}{\ \ \ \ \ \ \ }\tau_{{i}}=\omega t_{{i}}{\ \ }$%
\end{tabular}
\   \label{A13}
\end{equation}
$\gamma$, here, is the Keldysh parameter \cite{Keldysh65}. The minimum
physically meaningful value of $t_{\mathrm{i}}$ is found by setting the root
terms equal to zero in the above. Smaller values result in a complex $G$,
inconsistent with it being defined as real by Eq.~\ref{A08}.

\subsection{Most probable $G$}

Although prefactor $C$, like $G$, also depends on $t_{\mathrm{i}}$, it does
to a much lesser degree than the exponential term in Eq.~\ref{A09}, so we
base our most probable path on the extrema of $G$ alone. Fig.~A1a shows that
the critical (minimum) value of $G$, which we identify as $G=G_{\mathrm{c}}$
at $t_{\mathrm{i}}$= $t_{\mathrm{ic}}$, occurs at a local differential
extremum for $\varepsilon =1$. Figure~A1b, however, shows that the extremum
identifying these parameters is at beginning of the trace for $\varepsilon
=0 $. For high values of $N_{\mathrm{q}}=U_{0}/\left( \hbar \omega \right) $%
, one sees from these plots that $W$ drops rapidly as $t_{\mathrm{i}}$
departs from $t_{\mathrm{ic}}$. For O$_{2}$ ($U_{0}=12.063$ eV \cite%
{Samson66}) exposed to 800 nm ($\hbar \omega =1.550$ eV), we have $N_{%
\mathrm{q}}=7.78$. Each vertical unit, then, corresponds to a factor of $%
\exp \left( N_{\mathrm{q}}\right) =2527$ decrease in $W$.

$G_{\mathrm{c}}$ and $t_{\mathrm{ic}}$ vs.\ $t_{0}$ for both polarities are
plotted in Fig.~A2a and Fig.~A2b, respectively. $t_{\mathrm{ic}}$ is
determined numerically from the minimum of $G$ for $\varepsilon =1$, and
from $R_{3}=0$ in Eq.~\ref{A13}.7 (corresponding to the beginning of $G$
being real) for $\varepsilon =0$. We have, then, for the latter,

\begin{equation}
\begin{tabular}{l}
$\varepsilon=0$:\ \ $\frac{2G_{\mathrm{c}}}{\hbar}=\frac{U_{0}}{\hbar\omega }%
\left( \left( \frac{1+2\sin^{2}\tau_{0}}{\gamma^{2}}+\frac{\allowbreak 2}{%
\cos^{2}\tau_{0}}\right) \tau_{\mathrm{ic}}\right. $ \\ 
$\left. -\frac{1+2\sin^{2}\tau_{0}\allowbreak}{\gamma\left\vert \cos\tau
_{0}\right\vert }\sqrt{1+\frac{\gamma^{2}}{\cos^{2}\tau_{0}}}\right) $ \\ 
$\tau_{\mathrm{ic}}=\ln\left( \frac{\gamma}{\left\vert \cos\tau
_{0}\right\vert }+\sqrt{1+\frac{\gamma^{2}}{\cos^{2}\tau_{0}}}\right) $%
\end{tabular}
\   \label{A14}
\end{equation}

\begin{figure}[H]\centering\includegraphics[width=3.375in]{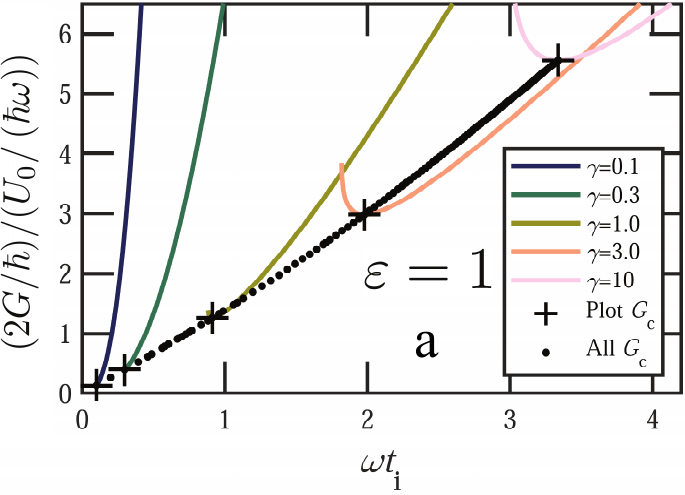}\end{figure}     
\begin{figure}[H]\centering\includegraphics[width=3.375in]{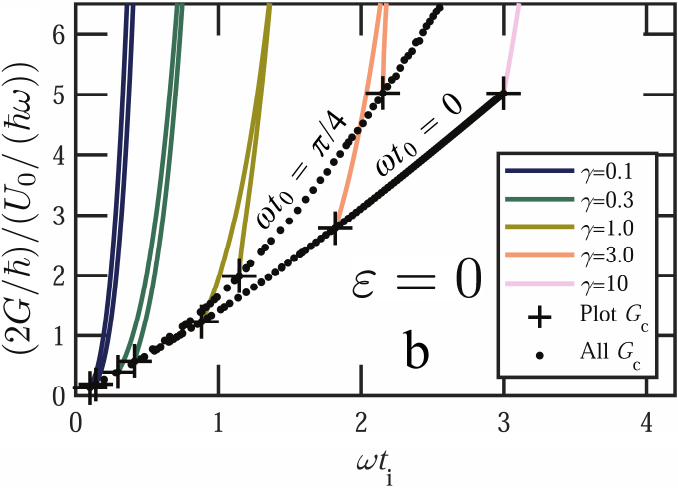}\end{figure}     
\textbf{Fig.~A1} Ionization rate negative exponential term $G$ vs.\ tunnel
time $t_{\mathrm{i}}$ (both normalized to span parameter space) for (a) $%
\varepsilon=1$ and (b) $\varepsilon=0$ . Plots are for a range of $\gamma$
values. The dots forming a line mark the critical (minimum) value of $G=G_{%
\mathrm{c}}$ for $G$ curves (plotted or not) with a $\gamma$ increment of $%
0.1$. $G$ is also a function of ionization time $t_{0}$ for $\varepsilon =0$%
. It is plotted for two $t_{0}$ values in (b) to show trending.

\subsection{Most probable post-optical and residual momenta}

The most probable value of $\mathbf{p}_{{\mathrm{f}r}}$ is found by
substituting $\tau_{\mathrm{ic}}$ into the expression for $\xi$ of Eq.~\ref%
{A10}.3, and into the $\eta$ expressions in Eqs.~\ref{A12}. $\xi$ and $\eta$
so determined are then substituted into Eq.~\ref{A11}.3 and Eq.~\ref{A11}.5.
The results are,%
\begin{equation}
\begin{tabular}{l}
$\varepsilon=1$:\ $p_{{\mathrm{f}x}}=p_{{\mathrm{f}r}}\sin\tau_{0}\ \ p_{{%
\mathrm{f}y}}=-p_{{\mathrm{f}r}}\cos\tau_{0}$ \\ 
$p_{{\mathrm{f}r}}=\frac{\sqrt{2mU_{0}}}{\gamma}\left( \cosh\tau _{\mathrm{ic%
}}-\sqrt{\sinh^{2}\tau_{\mathrm{ic}}-\gamma^{2}}\right) $ \\ 
$\varepsilon=0$:\ $p_{{\mathrm{f}x}}=\frac{\sqrt{2mU_{0}}\sin\tau_{0}}{%
\gamma }\allowbreak\sqrt{1+\frac{\gamma^{2}}{\cos^{2}\tau_{0}}}$ \ $p_{{%
\mathrm{f}y}}=0$%
\end{tabular}
\label{A15}
\end{equation}
We see from this and Eq.~\ref{A01}.3 that $\mathbf{p}_{{\mathrm{f}r}}$ is
perpendicular to $\mathbf{E}$ for $\varepsilon=1$, and $p_{{\mathrm{f}r}}=p_{%
{\mathrm{f}x}}$ for $\varepsilon=0$. $p_{{\mathrm{f}r}}$ is plotted in
Fig.~A3a for both polarities.

\begin{figure}[H]\centering\includegraphics[width=3.375in]{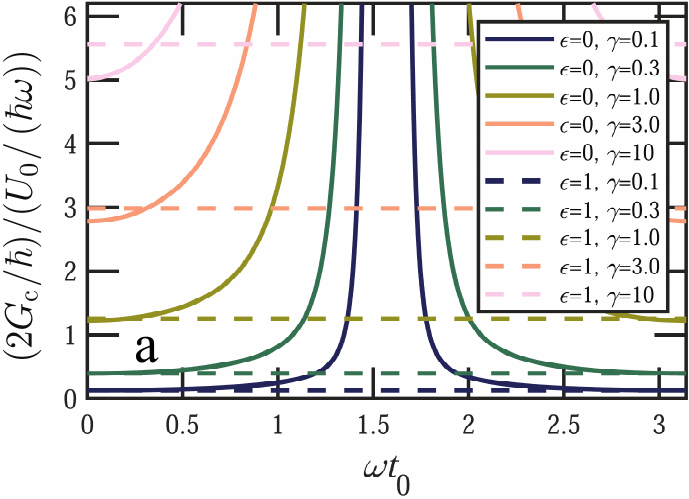}\end{figure}     
\begin{figure}[H]\centering\includegraphics[width=3.375in]{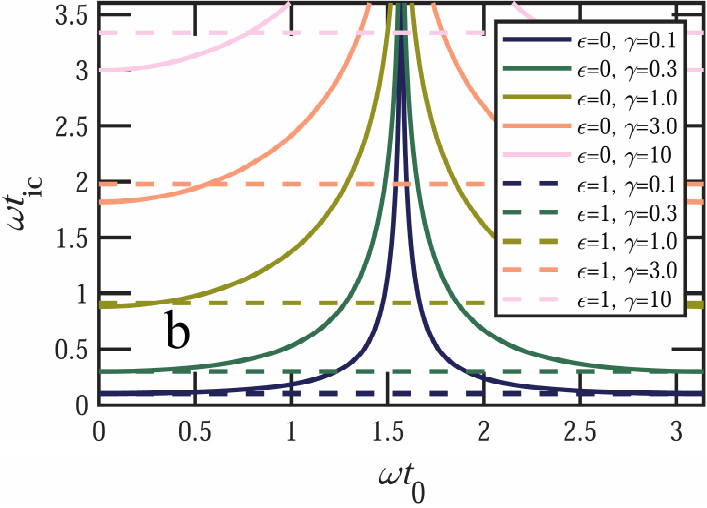}\end{figure}     
\textbf{Fig.~A2} Critical (most probable) values of $G$ (defined as $G_{%
\mathrm{c}}$) and their tunnel time $t_{\mathrm{i}}=t_{\mathrm{ic}}$ vs.\
ionization time $t_{0}$ are plotted for a range of $\gamma $ values. Plots
for $\varepsilon =1$ are flat lines since $G_{\mathrm{c}}$ does not vary
with $t_{0}$ for circular polarization.

\bigskip

By solving Eq.~\ref{A04} for $\mathbf{p}_{{\mathrm{f}r}}$ at $t=t_{0}$ and $%
z=0$, we have an independent expression for this term, 
\begin{equation}
\mathbf{p}_{{\mathrm{f}r}}=\mathbf{p}_{0r}-e\mathbf{A}_{0}\left(
t_{0}\right) \ \ \   \label{A16}
\end{equation}
where $\mathbf{p}_{0r}$ (referred to as \textquotedblleft residual
momentum\textquotedblright) is the electron momentum immediately upon
ionization at $t=t_{0}$. From Eq.~\ref{A01}.1, then, at $t=t_{0}$ and $z=0$,

\begin{equation}
\begin{tabular}{l}
$p_{{\mathrm{f}x}}=p_{0x}+\frac{e\mathcal{E}}{\omega}\sin\omega t_{0}$ \\ 
$p_{{\mathrm{f}y}}=p_{0y}-\frac{\varepsilon e\mathcal{E}}{\omega}\cos\omega
t_{0}$%
\end{tabular}
\   \label{A17}
\end{equation}
$\mathbf{p}_{0r}$ components are determined by equating these expressions to
those in Eqs.~\ref{A15} and solving for them. The result is,

\begin{equation}
\begin{tabular}{l}
$\varepsilon=1$:\ \ $p_{0x}=p_{0r}\sin\tau_{0}$ \ \ \ \ $p_{0y}=p_{0r}\cos
\tau_{0}$ \\ 
$p_{0r}=\frac{\sqrt{2mU_{0}}}{\gamma}\left( \cosh\tau_{\mathrm{ic}}-\sqrt{%
\sinh^{2}\tau_{\mathrm{ic}}-\gamma^{2}}-1\right) $ \\ 
$\varepsilon=0$:\ \ $p_{0x}=\frac{\sqrt{2mU_{0}}}{\gamma}$ \\ 
$\times\left( \sin\tau_{0}\sqrt{1+\frac{\gamma^{2}}{\cos^{2}\tau_{0}}}%
-\sin\tau_{0}\right) \ \ \ \ \ p_{0y}=0$%
\end{tabular}
\   \label{A18}
\end{equation}
$\ \mathbf{p}_{0r}$, like $\mathbf{p}_{{\mathrm{f}r}}$, is perpendicular to $%
\mathbf{E}$ for $\varepsilon=1$. The results are plotted in Fig.~A3a.

An expression for $p_{{\mathrm{f}z}}$ is found as a second order
contribution to $\mathbf{p}_{\text{f}}$ by\ using the first order
approximation to $\mathbf{p}_{r}$ (Eq.~\ref{A04}) in the r.h.s.\ of Eq.~\ref%
{A03}.2 (with $\partial \phi /\partial \mathbf{x}$ still not yet considered)
to obtain, for both polarities, 
\begin{equation}
\frac{\partial p_{z}}{\partial t}\mathbf{\hat{e}}_{z}=-\frac{1}{m}\mathbf{p}%
_{r}\times \nabla \times \mathbf{p}_{r}\mathbf{=-}\frac{1}{2m}\frac{\partial 
}{\partial z}p_{r}^{2}\mathbf{\hat{e}}_{z}  \label{A19}
\end{equation}%
We approximate here $d/dt$ by $\partial /\partial t$ (omitting the
convective term) since the electron, being nonrelativistic, has a
displacement over an optical cycle much smaller than an optical wavelength.
\ We have used $\nabla \times \mathbf{p}_{r}=e\nabla \times \mathbf{A}$,
where $\mathbf{p}_{{\mathrm{f}r}}$ in Eq.~\ref{A04} does not contribute to
the curl since, as post-optical property, it is a constant of motion. The
final expression is all that is left of the second given that $\mathbf{p}%
_{r} $ has (by definition) no $z$ component and only a $z$ spatial
dependence.

The spatiotemporal dependence of $\mathbf{p}$ depends only on comoving
spatial coordinate $z^{\prime}=z-ct$ (referred to as being \textquotedblleft
steady state\textquotedblright) since it results from an optical wave with
that dependence. This means that at $z=0$ ($z^{\prime}=-ct$), we may
substitute $\partial/\partial z\rightarrow-c^{-1}\partial/\partial t$ in Eq.~%
\ref{A19}, resulting in, 
\begin{equation}
\frac{\partial}{\partial t}p_{z}\mathbf{=}\frac{1}{2mc}\frac{\partial }{%
\partial t}p_{r}^{2}  \label{A20}
\end{equation}
Integrating from $t=t_{0}$ to variable $t$,%
\begin{equation}
p_{z}-p_{0z}=\frac{p_{r}^{2}-p_{0r}^{2}}{2mc}  \label{A21}
\end{equation}

To determine $p_{0z}$, we first specify Eq.~\ref{A21} at $t=t_{\mathrm{s}}$,%
\begin{equation}
p_{z}\left( t_{\mathrm{s}}\right) -p_{0z}=\frac{p_{r}^{2}\left( t_{\mathrm{s}%
}\right) -p_{0r}^{2}}{2mc}  \label{A22}
\end{equation}
where $p_{z}\left( t_{\mathrm{s}}\right) $ and $p_{r}\left( t_{\mathrm{s}%
}\right) $ arguments are the momentum components at $t=t_{\mathrm{s}}$. From
Eq.~\ref{A04}, we see $p_{r}\left( t_{\mathrm{s}}\right) $ is equal to the
term in square brackets of (saddle) Eq.~\ref{A07}, implying, $%
p_{r}^{2}\left( t_{\mathrm{s}}\right) =-2mU_{0}$ and, therefore, from Eq.~%
\ref{A22},%
\begin{equation}
p_{0z}=p_{z}\left( t_{\mathrm{s}}\right) +\frac{U_{0}}{c}+\frac{p_{0r}^{2}}{%
2c}  \label{A23}
\end{equation}
$p_{z}\left( t_{\mathrm{s}}\right) $, like $p_{r}\left( t_{\mathrm{s}%
}\right) $, is an initial condition for the tunnel path and, therefore, an
intrinsic property of the ground state independent of the EM field. It
follows, then, that $p_{z}\left( t_{\mathrm{s}}\right) =-U_{0}/c$\ since
this ensures that as one drops the magnitude of $\mathcal{E}$ to the point
where $p_{\mathrm{{0}r}}\rightarrow0$, then $p_{\mathrm{{0}z}}\rightarrow0$
too. Given this,\ from Eq.~\ref{A23} and Eq.~\ref{A21}, respectively, 
\begin{equation}
p_{0z}=\frac{p_{0r}^{2}}{2mc}\ \ \ \ \ \ \ p_{{\mathrm{f}z}}=\frac {p_{{%
\mathrm{f}r}}^{2}}{2mc}\   \label{A24}
\end{equation}
where the latter has been solved for solved for $p_{z}=p_{{\mathrm{f}z}}$.
This generalizes Zhou's relationship \cite{Zhou11} between $p_{{\mathrm{f}z}%
} $ and $p_{{\mathrm{f}r}}$, for which Zhou assumes $\mathbf{p}_{0}=\mathbf{0%
}$. The results are plotted in Fig.~A3b.

\begin{figure}[H]\centering\includegraphics[width=3.375in]{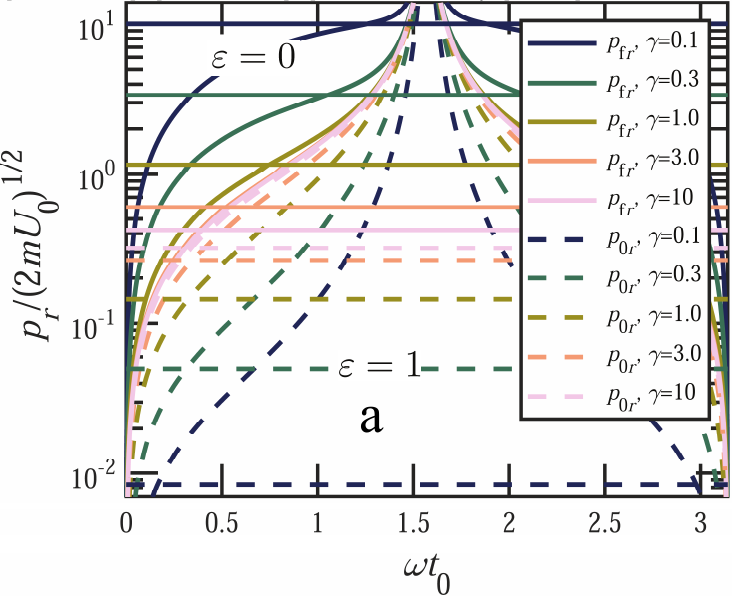}\end{figure}     
\begin{figure}[H]\centering\includegraphics[width=3.375in]{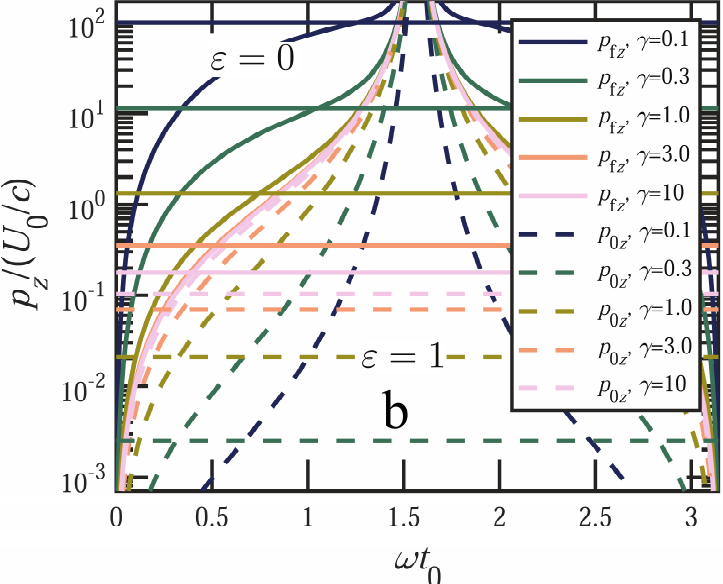}\end{figure}     
\textbf{Fig.~A3} Radial (a) and axial (b) components of post-optical (solid
lines) and residual (dashed lines) momenta $\mathbf{p}_{{\mathrm{f}}}$ and $%
\mathbf{p}_{0}$, respectively, vs.\ ionization time $t_{0}$ for the most
probable ionization path are plotted for a range of $\gamma $ values. The
radial momenta are perpendicular to $\mathbf{E}$ for $\varepsilon =1$ (flat
lines), and parallel for $\varepsilon =0$ (curved lines).

\subsection{$t_{0}$ dependent cofactor $C_{0}$}

We implement Li's solution \cite{Li17} to $C$'s $t_{0}$ dependent term $%
C_{0} $ due to the Coulomb interaction between the tunneling electron and
the charge distribution left behind as another second order correction to
the most probable tunnel path. The complex dynamic position vector
transverse to the $z$-axis $\mathbf{x}_{r}$ along the tunnel path for an
atom at $z=0$ unperturbed by the Coulomb potential $\phi $ is used to
calculate $C_{0}$.

$\mathbf{x}_{r}$ is found by integrating Eq.~\ref{A04} from $t_{\mathrm{s}}$
(when the electron is at the nucleus) to variable time $t$ over the tunnel
path, 
\begin{equation}
\begin{tabular}{l}
$m\mathbf{x}_{r}=\int\limits_{t_{\mathrm{s}}}^{t}\mathbf{p}_{r}\left(
t^{\prime \prime }\right) dt^{\prime \prime }$ \\ 
$\mathbf{p}_{r}\left( t^{\prime \prime }\right) =\mathbf{p}_{{\mathrm{f}r}}+e%
\mathbf{A}_{0}\left( t^{\prime \prime }\right) $%
\end{tabular}%
\ \ \ \ \ \ \   \label{A25}
\end{equation}%
From Eq.~\ref{A15} and Eq.~\ref{A01}.1 at $z=0$, and $\omega t=\tau
_{0}+i\tau ^{\prime }$ for the upper integration limit of the above, the
analytic integrations are performed and separated into real and imaginary
components. The components of $\mathbf{x}_{r}$ along the tunnel path ($0\leq
\tau ^{\prime }\leq \tau _{\mathrm{ic}}$) are,%
\begin{equation}
\begin{tabular}{l}
$\varepsilon =1$:\ \ $x_{r}=-r_{10}\left( \tau ^{\prime }\right) \cos \tau
_{0}-ir_{11}\left( \tau ^{\prime }\right) \sin \tau _{0}$ \\ 
$y_{r}=-r_{10}\left( \tau ^{\prime }\right) \sin \tau _{0}+ir_{11}\left(
\tau ^{\prime }\right) \cos \tau _{0}$ \\ 
$r_{10}\left( \tau ^{\prime }\right) =\sqrt{\frac{2U_{0}}{m\omega ^{2}}}%
\frac{1}{\gamma }\left( \cosh \tau _{\mathrm{ic}}-\cosh \tau ^{\prime
}\right) $ \\ 
$r_{11}\left( \tau ^{\prime }\right) =\sqrt{\frac{2U_{0}}{m\omega ^{2}}}%
\frac{1}{\gamma }$ \\ 
$\times \left( \left( \cosh \tau _{\mathrm{ic}}-\sqrt{\sinh ^{2}\tau _{%
\mathrm{ic}}-\gamma ^{2}}\right) \left( \tau _{\mathrm{ic}}-\tau ^{\prime
}\right) \right. $ \\ 
$\ \ \left. -\left( \sinh \tau _{\mathrm{ic}}-\sinh \tau ^{\prime }\right)
\right) $ \\ 
$\varepsilon =0$:\ \ $x_{r}=-r_{00}\left( \tau ^{\prime }\right)
-ir_{01}\left( \tau ^{\prime }\right) $ \\ 
$r_{00}\left( \tau ^{\prime }\right) =\sqrt{\frac{2U_{0}}{m\omega ^{2}}}%
\frac{\cos \tau _{0}}{\gamma }\allowbreak \left( \sqrt{1+\frac{\gamma ^{2}}{%
\cos ^{2}\tau _{0}}}-\cosh \tau ^{\prime }\right) $ \\ 
$r_{01}\left( \tau ^{\prime }\right) =-\sqrt{\frac{2U_{0}}{m\omega ^{2}}}%
\frac{\sin \tau _{0}}{\gamma }\left( \left( \frac{\gamma }{\left\vert \cos
\tau _{0}\right\vert }-\sinh \tau ^{\prime }\right) \right. $ \\ 
$\left. -\allowbreak \sqrt{1+\frac{\gamma ^{2}}{\cos ^{2}\tau _{0}}}\left(
\tau _{\mathrm{ic}}-\tau ^{\prime }\right) \right) $%
\end{tabular}
\label{A26}
\end{equation}%
Here, we have used the definition of $\gamma $ (Eq.~\ref{A13}.8) to
eliminate $\mathcal{E}$, and substituted the analytic expression for $\tau _{%
\mathrm{ic}}$ from Eq.~\ref{A14}.3 for $\varepsilon =0$ for most
occurrences. The signs of Eq.~\ref{A26}.7 and Eq.~\ref{A18}.3 imply that at $%
\tau ^{\prime }=0$ ($\tau =\tau _{0}$) for $\varepsilon =0$, the electron's
momentum is \emph{toward} the atom that it just tunneled out from for odd
optical quarter-cycles. Consider, though, that velocity and displacement of
a particle accelerated by an oscillating force are generally out of phase.

The Coulomb singularity at $\mathbf{x}_{r}=\mathbf{0}$ is compensated for in
Li's expression for $C_{0}$ \cite{Li17} by matching it with a term based on
the asymptotic wave function of the ground state. In our notation, the
result is, 
\begin{equation}
\begin{tabular}{l}
$C_{0}=\left( N_{\mathrm{q}}\tau _{\mathrm{ic}}\exp \left(
\int\limits_{0}^{\tau _{\mathrm{ic}}}\left( \sqrt{\frac{2U_{0}}{m\omega ^{2}}%
}\frac{\left\vert r_{0}\right\vert }{r_{0}^{2}+r_{1}^{2}}\right. \right.
\right. $ \\ 
$\left. \left. \left. -\frac{1}{\tau _{\mathrm{ic}}-\tau ^{\prime }}\right)
d\tau ^{\prime }\right) \right) ^{\frac{2Z}{\kappa }}$ \\ 
$N_{\mathrm{q}}=\frac{U_{0}}{\hbar \omega }$ $\ \ \ \ \kappa =\frac{4\pi
\epsilon _{0}\hbar }{e^{2}}\sqrt{\frac{2U_{0}}{m}}$ \ \ \ \ \ \ \ $Z=1$%
\end{tabular}
\label{A27}
\end{equation}%
where $Z=1$ (ionization level) in our case. $r_{0}$ and $r_{1}$ here are the
real and imaginary components of the dynamic electron radius $r_{r}$ during
tunneling, respectively, where, in terms of the coordinates defined in Eqs.~%
\ref{A26}, $r_{r}^{2}=x_{r}^{2}\left( \tau ^{\prime }\right)
+y_{r}^{2}\left( \tau ^{\prime }\right) $ \cite{Perelomov67}. The integral
is over the complex tunnel path, but the integration variable has been
changed to $\tau ^{\prime }$ so that all terms are real, for computational
purposes. From Eqs.~\ref{A26},

\begin{equation}
\begin{tabular}{l}
$\varepsilon=1$:\ \ $r_{0}=\sqrt{r_{10}^{2}\left( \tau^{\prime}\right)
-r_{11}^{2}\left( \tau^{\prime}\right) }$ \ \ $r_{1}=0$ \\ 
$\varepsilon=0$:\ \ $\left\vert r_{0}\right\vert =\left\vert r_{00}\left(
\tau^{\prime}\right) \right\vert $ $\ \ \ \ \ r_{1}^{2}=r_{01}^{2}\left(
\tau^{\prime}\right) $%
\end{tabular}
\   \label{A28}
\end{equation}
Fortunately, $r_{r}^{2}$ is real for both polarities, so the general
solution to the square root of a complex number \cite{Uspensky48} is not
needed (as it would be for other values of $\varepsilon$). Note that the
root term in the integrand of Eq.~\ref{A27}.1 cancels the equivalent term in
the $r_{r}$ expressions of Eqs.~\ref{A26}, so does not represent an extra
degree of freedom (like $N_{\mathrm{q}}$ and $\kappa$). The results are
plotted in Fig.~A4.

\begin{figure}[H]\centering\includegraphics[width=3.375in]{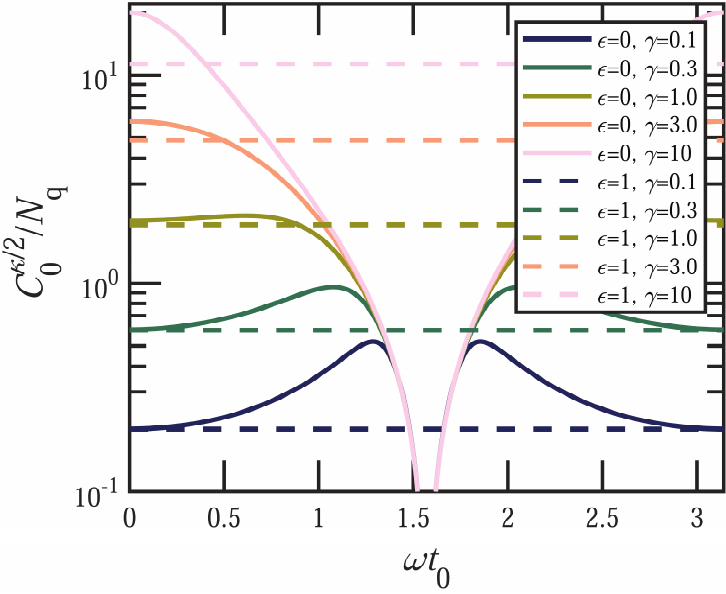}\end{figure}     
\textbf{Fig.~A4 } $C_{0}$ vs.\ $t_{0}$ for a range of $\gamma$ for our two
polarizations.

\subsection{Comparison to Luo's most probable path method}

The method for determining the most probable ionization path in this paper
is a two-step process. The saddle equation (Eq.~\ref{A07}) provides \emph{two%
} equations (its real and imaginary parts) that must be satisfied to
constrain \emph{three} dynamic variables: $p_{{\mathrm{f}x}}$, $p_{{\mathrm{f%
}y}}$, and $t_{\mathrm{i}}$. These equations are used to derive $p_{{\mathrm{%
f}x}}\ $and $p_{{\mathrm{f}y}}$ that satisfy the saddle equation as
functions of $t_{\mathrm{i}}$ and, upon substitution into Eq.~\ref{A08},
limit $G$ to be a function of $t_{\mathrm{i}}$ alone (Eqs.~\ref{A13}).
Finding the minimum $G_{\mathrm{c}}$ of\ $G$ w.r.t.\ $t_{\mathrm{i}}$ then
determines $t_{\mathrm{ic}}$ as the most probable $t_{\mathrm{i}}$. $G_{%
\mathrm{c}}$ is found to be the solution to $dG/dt_{\mathrm{i}}=0$ for $%
\varepsilon =1$\ and the smallest value of $t_{\mathrm{i}}$ for which $G$ is
real for $\varepsilon=0$. The aforementioned solution to $p_{{\mathrm{f}x}}$
and $p_{{\mathrm{f}y}}$ in terms of $t_{\mathrm{i}}=$ $t_{\mathrm{ic}}$ is
then used to find \emph{their} most probably values. The external equation
of motion (Eq.~\ref{A16}) is then used to determine the most probable $%
p_{0x} $ and $p_{0y}$.

The above approach differs from Luo's derivation \cite{Luo19}, where its
expression for $G$ (corresponding to our Eq.~\ref{A08}), \emph{unrestricted }%
by the saddle equation (our Eq.~\ref{A07}) is expressed as a function of $t_{%
\mathrm{i}}$ and residual momentum components $p_{0\Vert }$ and $p_{0\bot }$
parallel and perpendicular to optical $\mathbf{E}$, respectively. This is
accomplished by changing variables $p_{{\mathrm{f}x}}$ and $p_{{\mathrm{f}y}%
} $ (to which $G$ is originally a function of) to $p_{0\bot }$ and $%
p_{0\Vert } $, based on the external equation of motion (our Eq.~\ref{A16}).
Defining this $G$ here as $G_{\mathrm{L}}=G_{\mathrm{L}}\left( p_{0\bot
},p_{0\Vert },t_{\mathrm{i}}\right) $, Luo \cite{Luo19} takes $\partial G_{%
\mathrm{L}}/\partial p_{0\bot }=0$ solved for $p_{0\bot }$ as the most
probable $p_{0\bot }$, then equates this to an expression for $p_{0\bot }$
that \emph{is} restricted by the saddle equation, and solves for $t_{\mathrm{%
i}}$. This, then, is taken to be $t_{\mathrm{ic}}$ (the most probable $t_{%
\mathrm{i}}$), with the rest of the components of the most probable residual
and post-optical momenta following from that. Lou's method appears to result
in the same most probable path for\ $\varepsilon =1$, as inferred from
numerical trial solutions for $\varepsilon =1$, based on Luo's Eq.~21 \cite%
{Luo19} expression for minimizing $G$ for $\varepsilon =1$. The equivalence
is more definitive for $\varepsilon =0$ since Luo's Eq.~20 \cite{Luo19} and
critical $t_{\mathrm{i}}$ expression for $\varepsilon =0$ are analytically
equivalent to our Eqs.~\ref{A14}, provided parameter $a$ in the former's
root term is squared (an apparent typo). Equivalence of our different
methods has not been checked for $0<\varepsilon <1$. Our method for
determining $t_{\mathrm{i}}$ is presented since the underlying reason for
the equivalence, at least for $\varepsilon =0$ and $\varepsilon =1$, is not
understood.

\bigskip

\textbf{Funding. \ }This material is based on work supported by Air Force
Office of Scientific Research award FA9550-19RDCOR027.

\textbf{Acknowledgment. \ }The author thanks Jennifer A. Elle and Travis
Garrett for useful conversations.

\textbf{Disclosures. \ }The authors declare no conflicts of interest.

\textbf{Disclaimer.}\emph{\ \ } The views expressed are those of the author
and do not necessarily reflect the official policy or position of the
Department of the Air Force, the Department of Defense, or the U.S.
government.

\textbf{Data availability. \ }Data underlying the results presented in this
paper are available in the cited references.

\textbf{Release approval. \ }Approved for public release; distribution is
unlimited. Public Affairs release approval \#AFRL-2024-5092.

\bigskip

\textbf{Funding. \ }This material is based on work supported by Air Force
Office of Scientific Research award FA9550-19RDCOR027.

\textbf{Acknowledgment. \ }The author thanks Jennifer A. Elle and Travis
Garrett for useful conversations.

\textbf{Disclosures. \ }The authors declare no conflicts of interest.

\textbf{Disclaimer.}\emph{\ \ } The views expressed are those of the author
and do not necessarily reflect the official policy or position of the
Department of the Air Force, the Department of Defense, or the U.S.
government.

\textbf{Data availability. \ }Data underlying the results presented in this
paper are available in the cited references.

\textbf{Release approval. \ }Approved for public release; distribution is
unlimited. Public Affairs release approval \#AFRL-2024-5092.

\bigskip

\newif\ifabfull\abfulltrue

\end{document}